\documentclass[aps,pre,eqsecnum,amsmath,showpacs]{revtex4-1}

\usepackage{graphicx}
\usepackage{dcolumn}
\usepackage{bm}
\usepackage{color}

\usepackage{amssymb}

\def\a{{\alpha }}
\def\g{{\gamma }}
\def\b{{\beta }}
\def\z{{\zeta }}

\def\be{\begin{equation}}
\def\ee#1{\label{#1}\end{equation}}
\def\d{\textsf{d} }
\def\e{\textsf{e} }
\def\s{\textsf{s} }
\def\x{\textsf{x} }
 \def\bx{\mathbf{x} }
 \def\bp{\mathbf{p} }
 \def\p{\textsf{p} }
  \def\n{\textsf{n} }
 \def\pp{\textsf{P} }
 \def\q{\textsf{q} }
\def\k{\textsf{k} }
 \def\J{\textsf{J} }
\def\no{\nonumber}
\def\lb{\label}
\def\h{\textsf{h} }
\def\x{\textsf{x} }
\def\D{\textsf{D}}
\def\lb{\label}

\def\lab{{\langle\alpha\beta\rangle}}
\newcommand{\ben}{\begin{eqnarray}}
\newcommand{\een}{\end{eqnarray}}

\begin{document}
\title{Grad's moment method for relativistic gas mixtures of Maxwellian particles}
\author{Gilberto  M. Kremer}
\email{kremer@fisica.ufpr.br}
\affiliation{Departamento de F\'{\i}sica, Universidade Federal do Paran\'a, Curitiba, Brazil }
\author{Wilson Marques Jr.}
\email{marques@fisica.ufpr.br}
\affiliation{Departamento de F\'{\i}sica, Universidade Federal do Paran\'a, Curitiba, Brazil}

 \begin{abstract}
Mixtures of relativistic gases are analyzed within the framework of Boltzmann equation by using Grad's moment method. A relativistic mixture of $r$ constituent  is characterized by the moments of the distribution function: particle four-flows, energy-momentum tensors and third-order moment tensors. By using Eckart's decomposition and introducing $13r+1$ scalar fields -- related with the four-velocity, temperature of the mixture, particle number densities, diffusion fluxes, non-equilibrium pressures, heat fluxes and pressure deviator tensors -- Grad's distribution functions are obtained. Grad's distribution functions are used to determine the third-order tensors and their production terms for mixtures whose constituent's rest masses are not too disparate, so that it follows a system of  $13r+1$ scalar field equations. By restricting to a binary mixture characterized by the six fields of partial particle number densities, four-velocity and temperature, the remainder 21 scalar equations are used to determine the constitutive equations for the non-equilibrium pressures, diffusion fluxes, pressure deviator tensors and heat fluxes. Hence the Navier-Stokes and generalized Fourier and Fick laws are obtained and the transport coefficients of bulk and shear viscosities, thermal conductivity, diffusion, thermal-diffusion and diffusion-thermal are determined. Analytic expressions for the transport coefficients in the non-relativistic and ultra-relativistic limiting case are given. Furthermore, solutions of the relativistic field equations for the binary mixture are obtained in form of forced and free waves. In the low frequency limiting case the phase velocity and the attenuation coefficient are determined for forced waves. In the small wavenumber limiting case it is shown that there exist four longitudinal eigenmodes, two  of them corresponding to propagating sound modes and two associated with non-propagating diffusive modes.
\end{abstract}


\maketitle

 \section{Introduction}

We may state that the beginning of the relativistic kinetic theory goes back to 1911 when J\"uttner \cite{J1}  derived an equilibrium distribution function for a relativistic gas, which in the non-relativistic limiting case becomes the Maxwellian distribution function.  J\"uttner has also succeeded to derive in  1928 \cite{J2} the relativistic  Bose-Einstein and Fermi-Dirac  distribution functions. The covariant formulation of the Boltzmann equation was proposed by Lichnerowicz and Marrot \cite{LM} in the forties of the last century and in the sixties the determination of the transport coefficients from the Boltzmann equation by using the Chapman-Enskog methodology was obtained by Israel \cite{I} and Kelly \cite{K} .

 Mixtures of relativistic gases are important in the field  of astrophysics, in particular to  problems associated
with gases at high temperature  in a stellar interior. Within the framework of Boltzmann equation these mixtures were studied by several authors and among others we quote the works
\cite{Mar,St,VL,Her1,Her2,Gui,vLK,Kox1,Kox2,And,Groot,CK,KK}. In these works the usual methodology applied to determine the transport coefficients was the Chapman-Enskog method and general expressions for the transport coefficients were determined for generic differential cross sections. Only few explicit expressions for the transport coefficients were determined for very special cases of binary  mixtures of Maxwellian particles \cite{VL} and hard spheres \cite{Kox1} where the constituents have the same rest masses  or a Lorentz gas of hard spheres \cite{Kox2}.

In this work   we analyze mixtures of relativistic gases of Maxwellian particles by using Grad's moment method applied to the system of Boltzmann equations. In relativistic kinetic theory there is no unique differential cross section which tends in the non-relativistic limiting case to the one which is proportional to the relative velocity, known as the differential cross section of Maxwellian particles. At least three of them given in the works \cite{I,Po,GK} have this property, and here we shall use the one defined in \cite{GK}.

We are interested in a mixture that is characterized by $r$ constituents and described by $13r+1$ basic fields of  four-velocity, temperature of the mixture, particle number densities, diffusion fluxes, non-equilibrium pressures, heat fluxes and pressure deviator tensors.  Grad's distribution functions are obtained from the definition of the basic fields which are related with the Eckart decomposition of the particle four-flows and of the energy-momentum tensors. From the knowledge of Grad's distribution functions the third-order moments of the distribution functions are calculated and the  production terms that appear in the balance equations are determined for the case of a mixture of Maxwellian particles for constituents whose rest masses are not too disparate. Once the $13r+1$ field equations for the basic fields are established, the mixture is restricted to a binary mixture characterized by six fields of particle number densities, four-velocity and temperature of the mixture. The constitutive equations for the diffusion flux, pressure deviator tensors, non-equilibrium tensors and heat fluxes are determined by a method akin to the Maxwellian iteration procedure applied to the remaining 21 scalar field equations. The generalized laws of Fick, Fourier and Navier-Stokes are obtained and the coefficients of diffusion, thermal-diffusion, diffusion-thermal, thermal conductivity and bulk and shear viscosities are determined. Explicit expressions for these coefficients in the non-relativistic and ultra-relativistic limiting cases are given, as well as their graphics as functions of a parameter which represents the ratio of the rest energy of a particle and the thermal energy of the gas. From  the system of field equations for the binary mixture we have analyzed the solutions corresponding to small perturbations from an equilibrium state  related with the propagation of forced and free waves.  For the acoustic solution in the low frequency limiting case we have determined the phase velocity and the attenuation coefficient. For the eigenmodes in the small wavenumber limiting case we have shown the occurrence of two propagating sound modes and two non-propagating diffusive modes.

This work is structured as follows: in Section II  the Boltzmann equations, the moments of the distribution functions and their balance equations are introduced. The Eckart decomposition is the subject of Section III and Grad distribution functions are determined in Section IV. In Section V the constitutive equations for the third-order moment tensors and for the production terms are obtained and in Section VI the linearized $13r+1$ field equations are established. A binary mixture with six scalar fields of particle number densities, four-velocity and temperature of the mixture is analyzed in Section VII, where the generalized laws of Fick, Fourier and Navier-Stokes are obtained from a method akin to the Maxwellian iteration procedure. Furthermore, the transport coefficients of diffusion, thermal-diffusion, diffusion-thermal, thermal conductivity and bulk and shear viscosities associated with these laws are determined in the non-relativistic and ultra-relativistic limiting cases. In Section VIII  solutions of the binary mixture  field equations are analyzed in terms of sound propagation and eigenmodes. Finally, the main conclusions of this work are discussed in Section IX.

Latin indexes running from $a=1\dots r$ specify the constituents of the mixture, while Greek indexes running from $\alpha=0,1,2,3$ denote the space-time coordinates.

 \section{Boltzmann and Transfer Equations}

Let us consider a relativistic gas mixture of $r$  constituents in a Minkowski space with metric tensor $(\eta_{\a\b})={\rm diag}(1,-1,-1,-1)$. The particles of constituent $a=1,\dots r$ have rest mass $m_a$ and are characterized by the space-time coordinates $(x^\a)=(ct,\bx)$ and momenta  $(p_a^\a)=(p_a^0,\bp_a)$. The length of the momentum four-vector is $m_a c$ so that $p_a^0=\sqrt{\vert\bp_a\vert^2+m_a^2c^2}$.

An elastic collision between the particles of two constituent $a$ and $b$ whose momentum four-vectors before collision are denoted by $p_a^\a$ and $p_b^\a$ reads
\be
p_a^\alpha+p_b^\alpha=p'^\alpha_{a}+p'^\alpha_{b},
\ee{1}
where the primes denote the  momentum four-vectors after collision.

A state of the relativistic mixture of $r$ constituents is characterized by the set
of one-particle distribution functions
\be
f({\bf x},{\bf p}_a,t)\equiv f_a, \qquad
a=1,2,\dots,r
\ee{2}
such that $f({\bf x},{\bf p}_a,t) d^3x d^3p_a$
gives at time
$t$, the number of particles of constituent
$a$ in the volume element
$d^3 x$ about  ${\bf x}$ and with momenta in the range
$d^3p_a$ about ${\bf p}_a$.

The one-particle distribution function of constituent $a$ satisfies a Boltzmann
equation, which in the absence of external fields reads (see e.g. \cite{Groot,CK})
\be
p_a^\alpha \,\partial_\a f_a
=\sum_{b=1}^r\int(f_a'f_b'-f_a f_b)F_{ba}\sigma_{ab}
\,d\Omega \frac{d^3p_b}{p_{b0}},
\ee{3}
where $\sigma_{ab}$ and $d\Omega$
denote the invariant differential elastic
cross-section and the element of
solid angle that characterizes a
binary  collision
between the particles of constituent $a$ with those of constituent $b$,
respectively. Moreover,
$F_{ba}$ is the invariant  flux defined by
\be
F_{ba}={p^0_ap^0_b }\sqrt{
\left(\frac{\bp_a}{p_a^0} - \frac{\bp_b}{p_b^0}\right)^2-
\left(\frac{\bp_a}{p_a^0}\times \frac{\bp_b}{p_b^0}\right)^2}=\sqrt{(p^\a_ap_{b\a})^2-m_a^2m_b^2c^4},
\ee{4}
and the following abbreviations
were introduced
\be
f_a'\equiv f({\bf x},{\bf p}_a',t),\quad
f_b'\equiv f({\bf x},{\bf p}_b',t),\quad
f_a\equiv f({\bf x},{\bf p}_a,t),  \quad
f_b\equiv f({\bf x},{\bf p}_b,t).
\ee{5}

The general equation of transfer  for the
constituent  $a$ of the mixture
is obtained through
 the multiplication of the Boltzmann  equation
(\ref{3}) by an arbitrary function of the momentum four-vector $\psi_a\equiv\psi(p^\a_a)$ and
integration of the resulting equation over
all values of $d^3p_a/p_{a0}$, yielding:
\ben\no
{\partial_\alpha}\int
 \psi_a p_a^\alpha f _a
\frac{d^3 p_a}{p_{a0}}&=&\sum_{b=1}^r\int \psi_a (f_a'f_b'-f_a f_b)
 F_{ba}\sigma_{ab} d\Omega
\frac{d^3 p_b}{p_{b0}}
\frac{d^3 p_a}{p_{a0}}\\\lb{6}
&=&\sum_{b=1}^r\int(\psi_a'-\psi_a)f_a f_b F_{ba}
\sigma_{ab} d\Omega
\frac{d^3 p_b}{p_{b0}}
\frac{d^3 p_a}{p_{a0}},
\een
where the last equality on  the right-hand side of the above equation follows by using the symmetry properties of the collision  term.

If we sum  (\ref{6}) over all constituents we get a
general equation of transfer for the mixture that
reads
\ben\lb{7}
{\partial_\alpha}\sum_{a=1}^r
\int \psi_a p_a^\alpha f_a \frac{d^3 p_a}{p_{a0}}=
\frac{1}{2}\sum_{a,b=1}^r\int (\psi_a'+\psi_b'-\psi_a-
\psi_b) f_a f_b F_{ba} \sigma_{ab} d\Omega
\frac{d^3 p_b}{p_{b0}}
\frac{d^3 p_a}{p_{a0}}.
\een
The   right-hand side of (\ref{7})
was also obtained by using the symmetry
properties of the collision  term and by  interchanging
the dummy indexes $a$ and $b$ in the sums.

The moments of the distribution  function we are
interested  are the partial particle four-flow  $N_a^\alpha$, the
partial energy-momentum  tensor
$T_a^{\alpha\beta}$ and the partial third-order moment tensor $T_a^{\alpha\beta\g}$ which are defined in terms of the distribution function through:
\be
N_a^\alpha=c\int p_a^\alpha f_a \frac{d^3 p_a}{p_{a0}},
\qquad T_a^{\alpha\beta}=c\int
p_a^\alpha p_a^\beta f_a
\frac{d^3 p_a}{p_{a0}},
\qquad T_a^{\alpha\beta\g}=c\int
p_a^\alpha p_a^\beta p_a^\g f_a
\frac{d^3 p_a}{p_{a0}}.
\ee{8}
The particle four-flow  of the mixture $N^\alpha$, the energy-momentum  tensor
$T^{\alpha\beta}$ and the third-order moment $T^{\alpha\beta\g}$ of the mixture
are obtained by summing the partial quantities over all constituents, i.e.
\be
N^\alpha=\sum_{a=1}^r N^\alpha_a,
\qquad
T^{\alpha\beta}=\sum_{a=1}^r T_a^{\alpha\beta},\qquad
T^{\alpha\beta\g}=\sum_{a=1}^r T_a^{\alpha\beta\g}.
\ee{9}

The balance  equations for the
partial particle four-flow, for the partial energy-momentum tensor and for the partial third-order moment tensor are obtained by choosing
$\psi_a= c$, $\psi_a=cp_a^\b$ and $\psi_a=cp_a^\b p_a^\g$ in (\ref{6}),
yielding
\be
\partial_\alpha N^\alpha_a=0,\qquad  \partial_\a T_a^{\a\b}=\pp_a^\b,\qquad \partial_\a T_a^{\a\b\g}=\pp_a^{\b\g},
\ee{10}
respectively. Above, the production terms $\pp_a^\b$ and $\pp_a^{\b\g}$ are given by
\ben\lb{10a}
\pp_a^\b=\sum_{b=1}^rc\int (p_a^{\prime\b}-p_a^\b)f_a f_b F_{ba}\sigma_{ab} d\Omega
\frac{d^3 p_b}{p_{b0}}\frac{d^3 p_a}{p_{a0}},\\\lb{10b}
\pp_a^{\b\g}=\sum_{b=1}^rc\int (p_a^{\prime\b}p_a^{\prime\g}-p_a^\b p_a^{\g})f_a f_b F_{ba}
\sigma_{ab} d\Omega\frac{d^3 p_b}{p_{b0}}\frac{d^3 p_a}{p_{a0}}.
\een

By summing (\ref{10}) over all constituents of the mixture
we get the balance equations for the particle
four-flow, energy-momentum tensor and third-order moment tensor of the mixture that read
\be
\partial_\alpha N^\alpha=0,\qquad \partial_\a T^{\a\b}=0,\qquad \partial_\a T^{\a\b\g}=\pp^{\b\g}=\sum_{a=1}^r \pp_a^{\b\g}.
\ee{11}
These are the balance equations used in the relativistic theory of extended thermodynamics of a single fluid (see \cite{Ru1}).

\section{Eckart decomposition}

For the decomposition of Eckart we introduce the four-velocity $U^\a$ such that $U^\a U_\a=c^2$ and the projector
\be
\Delta^{\a\b}=\eta^{\a\b}-\frac{1}{c^2}U^\a U^\b,\qquad\hbox{with}\qquad \Delta^{\a\b}U_\a=0.
\ee{12}
Hence, we write the partial particle four-flow as
\be
N_a^\alpha=\n_a U^\alpha+\J_a^\alpha,\quad\hbox{such that}\quad
\J_a^\alpha U_\alpha=0.
\ee{13}
Here $\n_a$ is the particle number  density
and $\J^\alpha_a$ the
diffusion
flux of the constituent $a$ in
the mixture. They are defined in terms of the following projections of the partial particle four-flow
\ben
\n_a=\frac{1}{c^2}N_a^\alpha U_\alpha,\qquad \J_a^\alpha=\Delta_\beta^\alpha N_a^\beta.
\een
From the definitions (\ref{8}a) and
(\ref{13}) we can represent the diffusion flux as
\be
\J_a^\alpha=\Delta_\beta^\alpha c\int p_a^\beta f_a\frac{d^3 p_a}{p_{a0}}.
\ee{14}
Further by summing   (\ref{13}) over all constituents
it follows
\be
\n=\sum_{a=1}^r \n_a,\qquad \sum_{a=1}^{r} \J_a^\alpha=0,
\ee{15}
which gives the particle number density of the mixture
as the sum of the partial
particle number  densities and shows -  due
to the  constraint
(\ref{15}b) - that there exist only $(r-1)$
partial diffusion fluxes that are linearly
independent for a mixture
of $r$ constituents.

The decomposition of the partial
energy-momentum  tensor in the
Eckart  description is written as (see e.g. \cite{StWa,Heb,KP})
\be
T_a^{\alpha\beta}=\p_a^{\langle\alpha\beta\rangle}-(\p_a+\varpi_a)
\Delta^{\alpha\beta}+\frac{1}{c^2}
U^\alpha\left(\q_a^\beta+\h_a \J_a^\beta\right)
+\frac{1}{c^2}
U^\beta\left(\q_a^\alpha+\h_a \J_a^\alpha\right)+\frac{\n_a\e_a}{c^2}U^\alpha
U^\beta,
\ee{16}
where we have introduced the partial quantities: $\p_a$   hydrostatic pressure, $\varpi_a$
non-equilibrium pressure, $\e_a$  energy  per particle,
$\q_a^\alpha$  heat  flux, $\h_a=\e_a+\p_a/\n_a$
enthalpy  per particle and $\p_a^{\langle\alpha\beta\rangle}$  pressure deviator tensor, i.e., the traceless part of the pressure tensor.
These fields are given in terms of the projections of the partial energy-momentum tensor by:
\ben
\e_a=\frac{1}{\n_a c^2}U_\alpha T_a^{\alpha\beta} U_\beta,\qquad \p_a+\varpi_a=-\frac{1}{3}\Delta_{\alpha\beta}T_a^{\alpha\beta},\qquad
\\
\q_a^\alpha+\h_a \J_a^\alpha=\Delta_\gamma^\alpha T_a^{\gamma\beta}U_\beta,\qquad
\p_a^{\langle\alpha\beta\rangle}=\left(\Delta^\alpha_\gamma\Delta^\beta_\delta-\frac{1}{3}\Delta^{\alpha\beta}\Delta_{\gamma\delta}\right)T_a^{\gamma\delta}.
\een
Note that the symmetric and traceless part of a second order tensor $\Upsilon^{\a\b}$ is defined in terms of the projector $\Delta^{\a\b}$ through the relationship
\ben\label{ups}
\Upsilon^{\langle\a\b\rangle}=\left[\frac{\left(\Delta^{\alpha}_{\gamma}
\Delta^{\beta}_{\delta}
+\Delta^{\alpha}_{\delta}
\Delta^{\beta}_{\gamma}\right)}{2}
-\frac{ \Delta^{\alpha\beta}\Delta_{\gamma\delta}}{3}
\right]\Upsilon^{\g\delta}.
\een

If we
sum the partial
energy-momentum  tensors over all constituents
of the mixture and compare
the resulting equation with the energy-momentum
tensor of the mixture
\be
T^{\alpha\beta}=\p^{\langle\alpha\beta\rangle}-(\p+\varpi)
\Delta^{\alpha\beta}+\frac{1}{c^2}
\left(U^\alpha \q^\beta+U^\beta \q^\alpha\right)+\frac{\n\e}{ c^2}U^\alpha U^\beta,
\ee{17}
we can identify the quantities of the mixture
\ben\lb{18}
\p^{\langle\alpha\beta\rangle}=\sum_{a=1}^r\p_a^{\langle\alpha\beta\rangle},
\qquad
\p=\sum_{a=1}^r \p_a,\qquad \varpi=\sum_{a=1}^r\varpi_a,\qquad \n\e=\sum_{a=1}^r
\n_a \e_a,\\\lb{18a} \n\h=\sum_{a=1}^r\n_a \h_a,\qquad
\q^\alpha=\sum_{a=1}^r\left(\q_a^\alpha+\h_a \J_a^\alpha\right).
\een
Note that according to (\ref{18a}b) the heat  flux of the
mixture is a sum of
the partial heat  fluxes and a term which represents the
transport of heat
due to diffusion.

\section{Grad's distribution function}

By supposing that all constituents are at the same temperature $T$ -- which represents the mixture's temperature -- we may describe a relativistic gas mixture by $13r+1$ basic scalar fields:
\ben\lb{fi}
\left\{
  \begin{array}{ll}
    \n_a, & \hbox{$r$ scalar fields;} \\
    U^\a, & \hbox{$3$ scalar fields;} \\
    \J_a^\a, & \hbox{$3(r-1)$ scalar fields;} \\
    T, & \hbox{$1$ scalar field;}\\
    \varpi_\a, &\hbox{$r$ scalar fields};\\
    \q_a^\a, &\hbox{$3r$ scalar fields};\\
    \p_a^{\langle\a\b\rangle}, &\hbox{$5r$ scalar fields}.
  \end{array}
\right.
\een

Unlike the non-relativistic theory of gases, the temperature in the relativistic theory cannot be defined in terms of the distribution function. It appears as the reciprocal of the integrating factor of the Pfaffian form $d\e-\p d\n/\n^2$, which identifies the potential of the Gibbs equation in equilibrium with the entropy per particle $\s$ such that $d \s =\frac{1}{T}\left(d\e-\frac{\p}{\n^2}d\n\right)$. Furthermore, in this work we are interested in process close to equilibrium where $\J_a^\alpha, \varpi_a, \q_a^\alpha$ and $\p_a^{\langle \alpha\beta\rangle}$ are considered as small quantities.

In terms of the $13r+1$ basic fields the  distribution function of constituent $a$ can be represented as the following polynomial function of the momentum four-vector:
\be
f_a=f_a^{(0)}\left(1+A_a^\a \,p_{a\a}+A_a^{\a\b}\,p_{a\a} \,p_{a\b}\right),
\ee{19}
where $A_a^{\a}$ and $A_a^{\a\b}$ are tensorial coefficients to be determined from the definitions of the partial particle four-flow (\ref{8}a) and partial energy-momentum tensor (\ref{8}b). Furthermore, $f_a^{(0)}$ is the Maxwell-J\"uttner distribution function
\be
f_a^{(0)}=\frac{\n_a}{4\pi kTm_a^2cK_2(\zeta_a)}\exp\left(-\frac{p_a^\a U_\a}{kT}\right).
\ee{20}
Above $k$ is the Boltzmann constant,  $K_2(\z_a)$ denotes the modified Bessel function of second kind defined through the integral
\ben
K_n(\zeta)=\left(\frac{\zeta}{2}\right)^n\frac{\Gamma(1/2)}{\Gamma(n+1/2)}\int_{1}^\infty e^{-\zeta y}\left(y^2-1\right)^{n-1/2}\,dy,
\een
and $\zeta_a=m_ac^2/kT$ is a parameter which represents the ratio of the rest energy $m_ac^2$ of a relativistic $a-$particle  and the thermal energy of the mixture $kT$. If $\zeta_a\gg1$ the constituent behaves as a non-relativistic gas, while when $\zeta_a\ll1$  as an ultra-relativistic gas.

In order to determine the tensorial coefficients we decompose the momentum four-vector $p_a^{\a}$ and the tensorial coefficients $A_a^{\a}$ and $A_a^{\a\b}$  as
\ben\label{22a}
p_a^\a=\Delta_\b^\a p_a^\b+\frac{1}{c^2}U^\a\left(p_a^\b U_\b\right),
\qquad
A_a^\a=\lambda_1 U^\a+\Lambda_1^\b\Delta_\b^\a,\\\label{22b}
A_a^{\a\b}=\lambda_2 U^\a U^\b+\lambda_3\eta^{\a\b}+\frac{1}{2}\Lambda_2^\g\left(\Delta_\g^\a U^\b+\Delta^\b_\g U^\a\right)+\Lambda^{\langle\g\delta\rangle}\Delta_\g^\a \Delta_\delta^\b.
\een
Here we have introduced new coefficients: the scalars $\lambda_1$, $\lambda_2$ and $\lambda_3$, the four-vectors $\Lambda_1^\a$ and $\Lambda_2^\a$ and the symmetric and traceless second-order tensor  $\Lambda^{\langle\g\delta\rangle}$.

Hence, the distribution function (\ref{19}) can be written as
\ben\no
f_a=f_a^{(0)}\Big(1+\lambda_1 p_a^\a U_\a+\lambda_2 p_a^\a p_a^\b U_\a U_\b+\lambda_3 m_a^2c^2+\Lambda_1^\b p_a^\a\Delta_{\a\b}
\\\lb{23}
+\Lambda_2^\g p_a^\a p_a^\b \Delta_{\g\a}U_\b+\Lambda^{\langle\g\delta\rangle}p_a^\a p_a^\b\Delta_{\a\g} \Delta_{\b\delta}\Big),
\een
thanks to  (\ref{22a}) and (\ref{22b}).

Now the insertion of (\ref{23}) into the definition of the partial particle four-flow (\ref{8}a) and of the partial energy-momentum tensor (\ref{8}b)  and integration of the resulting equations leads to the determination of the lambda tensorial coefficients and it follows Grad's distribution function for the constituent $a$, namely,
\ben\nonumber
f_a&=&f_a^{ (0)}\Biggl\{1-\frac{\J_{ a\a}}{\p_a}p_a^\a+
\frac{\varpi_a}{\p_a}
\frac{1-5G_a\zeta_a-\zeta_a^2+G_a^2\zeta_a^2}{
20 G_a+3\zeta_a-13 G_a^2\zeta_a-
2G_a \zeta_a^2+2G_a^3\zeta_a^2}
\Bigg[\frac{\zeta_a}{m_a^2 c^4}U_\a U_\b p^\a_a p^\b_a
\\\nonumber
&+&\frac{3\zeta_a}{m_a c^2}
\frac{6G_a+\zeta_a-G_a^2\zeta_a}{
1-5G_a\zeta_a-\zeta_a^2+G_a^2\zeta^2_a}
U_\a p_a^\a+
\frac{15G_a+2\zeta_a-6G_a^2\zeta_a+5G_a\zeta_a^2
+\zeta_a^3-G_a^2\zeta_a^3}{
1-5G_a\zeta_a-\zeta_a^2+G_a^2\zeta^2_a}
\Biggr]
\\\label{24}
&+&\frac{\q_{a\a}}{ \p_a}
\frac{\zeta_a}{ \zeta_a+5G_a-G_a^2\zeta_a}
\left[\frac{G_a}{ m_a c^2}p_a^\a-\frac{1}{ m_a^2 c^4}U_\b p_a^\a p_a^\b\right]
+\frac{\p_{ a\langle\a\b\rangle}}{ \p_a}
\frac{\zeta_a}{ 2G_a}\frac{1}{ m_a^2 c^2} p_a^\a p_a^\b
\Biggl\}.
\een
Above we have introduced the abbreviation $G_a=K_3(\z_a)/K_2(\z_a)$.

\section{Constitutive equations}

In order to determine the constitutive equation for the
third-order  moment tensor of constituent $a$
 we insert Grad\ distribution
function (\ref{24}) into its definition (\ref{8}c)
and integrate the resulting equation, yielding
\ben\nonumber
T_a^{\alpha\beta\g}=\left(\n_a C_{1a}+C_{2a}\varpi_a\right) U^{\alpha}U^{\beta}U^{\g}
+\frac{c^2}{ 6}\left(\n_am_a^2-\n_a C_{1a}-C_{2a}\varpi_a\right)
\left(\eta^{\alpha\beta} U^{\g}+\eta^{\alpha\g} U^{\beta}\right.
\\\nonumber
\left.+\eta^{\beta\g} U^{\alpha}\right)+C_{3a}\left(\eta^{\alpha\beta} \q_a^{\g}+\eta^{\alpha\g} \q_a^{\beta}+
\eta^{\beta\g} \q_a^{\alpha}\right)-\frac{6}{c^2}C_{3a}\left(U^{\alpha}
U^{\beta}\q_a^{\g}+U^{\alpha} U^{\g}\q_a^{\beta}
+U^{\beta} U^{\g}\q_a^{\alpha} \right)
\\\no
+C_{4a}\left(\p_a^{\langle\alpha\beta\rangle}U^{\g}+
\p_a^{\langle \alpha\g\rangle}U^{\beta}
+\p_a^{\langle \beta\g\rangle}U^{\alpha}\right)
+C_{5a}\left(\eta^{\alpha\beta} \J_a^{\g}+\eta^{\alpha\g} \J_a^{\beta}+
\eta^{\beta\g} \J_a^{\alpha}\right)\\\label{26}+\frac{C_{6a}}{c^2}\left(U^{\alpha}
U^{\beta}\J_a^{\g}+U^{\alpha} U^{\g}\J_a^{\beta}
+U^{\beta} U^{\g}\J_a^{\alpha} \right).
\een
The scalar coefficients $C_{1a}$ through $C_{6a}$ are given by
\ben\label{27}
C_{1a}=\frac{m_a^2}{\zeta_a}(\zeta_a+6G_a),\qquad C_{3a}=-\frac{m_a}{\zeta_a}\frac{\zeta_a+6G_a-G_a^2\zeta_a}{\zeta_a+5G_a-G_a^2\zeta_a},
\\
C_{2a}=-\frac{6m_a}{ c^2\zeta_a}\frac{2\zeta_a^3-5\zeta_a
+(19\zeta_a^2-30)G_a-(2\zeta_a^3-45\zeta_a)G_a^2-9\zeta_a^2 G_a^3}{20G_a
+3\zeta_a-13G_a^2\zeta_a-2\zeta_a^2G_a+2\zeta_a^2G_a^3},
\\\label{28}
\quad
C_{4a}=\frac{m_a}{G_a\zeta_a}(\zeta_a+6G_a),\quad C_{5a}=-m_akTG_a,\quad C_{6a}=m_akT\left(\zeta_a+6G_a\right).
\een

Furthermore, the pressure $\p_a$, the energy per particle $\e_a$ and the enthalpy per particle $\h_a$ of  constituent $a$ read
\be
\p_a=\n_a kT,\qquad \e_a=m_ac^2\left(G_a-\frac{1}{\zeta_a}\right),\qquad \h_a=m_ac^2G_a.
\ee{45}

For the determination of the production terms (\ref{10a}) and (\ref{10b}) it is
necessary to introduce the total momentum four-vector $P^\a$ and the
relative momentum four-vector $Q^\a$ defined by (see e.g. \cite{St})
 \ben\lb{28a}
 P^\a\equiv p_a^\a+p_b^\a,\qquad P^{\prime\a}\equiv p^{\prime\a}_a+p^{\prime\a}_b,
 \qquad
 Q^\a=p_a^\a-p_b^\a,\qquad
 Q^{\prime\a}=p^{\prime\a}_a-p^{\prime\a}_b.
 \een
   From the above equations together with the
 momentum four-vector conservation law it follows that
 \ben\lb{29}
 P^\a=P^{\prime\a},\qquad
 P^\a Q_\a=(m_a^2-m_b^2)c^2,
 \qquad
 Q^2=P^2-{2(m_a^2+m_b^2)c^2},
 \een
 where  $P^2=P^\a P_\a$ and $Q^2=-Q^\a Q_\a$ denote the magnitudes
 of the total and relative momentum four-vectors, respectively.
 The inverse transformations of (\ref{28a})   are
 \ben\lb{30a}
 p_a^\a=\frac{P^\a}{2}+\frac{Q^\a}{2},\qquad
 p_b^\a=\frac{P^\a}{2}-\frac{Q^\a}{2},
 \qquad
 p^{\prime\a}_a=\frac{P^\a}{2}+\frac{Q^{\prime\a}}{2},\qquad
 p^{\prime\a}_b=\frac{P^\a}{2}-\frac{Q^{\prime\a}}{2}.
 \een

The relative momentum four-vector can be written as
 \be
 Q^\a=\frac{(m_a^2-m_b^2)c^2}{ P^2} P^\a+\frac{\k^\a}{ P}\sqrt{P^4-2P^2(m_a^2+m_b^2)c^2+(m_a^2-m_b^2)^2c^4},
\ee{31}
where $\k^\a$ is a spacelike unit vector orthogonal to $P^\a$, i.e., $\k^\a P_\a=0$.

From now on we shall restrict ourselves to the case where the rest masses of the particles of the constituents are not too disparate so that $m_b\approx m_a(1+\epsilon),\, (\forall\, b=1,\dots, r\neq a)$ with $\epsilon$ denoting a small quantity. In this case we  have that $m_a^2-m_b^2\approx (m_a+m_b)\epsilon$ and the term  $(m_a^2-m_b^2)^2$ in (\ref{31}) can be neglected. In this case the relative momentum four-vector can be approximated by
 \be
 Q^\a=\frac{(m_a^2-m_b^2)c^2}{P^2} P^\a+Q\,{\k^\a}.
\ee{31a}
and the invariant flux (\ref{4}) reduces to
\be
F_{ba}=\sqrt{(p^\a_ap_{b\a})^2-m_a^2m_b^2c^4}=\frac{PQ}{2}.
\ee{32a}

For the determination of the production terms (\ref{10a}) and (\ref{10b}) we follow \cite{GK,CK} and introduce
the  invariant differential elastic cross-section for Maxwellian particles which can be written as
 \be
 \sigma_{ab}=\frac{P} {2\,c\,Q}\mathcal{F}(\Theta),
 \ee{33}
 where $\mathcal{F}(\Theta)$ is an arbitrary function of the scattering angle $\Theta$. By taking into account the expression (\ref{32a}) and (\ref{33})  we obtain that
 \be
   F_{ba}\sigma_{ab}=\frac{P^2}{4c}\mathcal{F}(\Theta).
 \ee{34}
This methodology is  similar to the one employed in the non-relativistic case and the simple expression (\ref{34}), which does not depend on the relative momentum four-vector, will permit us to integrate the production terms without the use of Grad's distribution function.

The calculation of the production terms is schematized in the Appendix A. Here we list only the final results
\ben\lb{35c}
&&\pp_a^\b=-\frac{\pi}{2c^2}\mathcal{B}_1\sum_{b=1}^r\left[\left(T^{\b\a}_a-T^\g_{a\g}\eta^{\b\a}\right)N_{b\a}
-\left(T^{\b\a}_b-T^\g_{b\g}\eta^{\b\a}\right)N_{a\a}\right],
\\\nonumber
&&\pp_a^{\a\b}=-\frac{\pi}{2c^2}\mathcal{B}_1\sum_{b=1}^r\Bigg\{T^{\a\b\g}_a N_{b\g}-T^{\a\b\g}_b N_{a\g}+
T^{\a\b}_a T^\g_{b\g}-T^{\a\b}_b T^\g_{a\g}\\\no
&&-\frac{\left(m_a^2-m_b^2\right)c^2}{2}\left(N_a^\a N_b^\b+N_a^\b N_b^\a\right)\Bigg\}
+\frac{\pi}{8c^2}\mathcal{B}_2\sum_{b=1}^r\Bigg\{\left(\frac{m_a^2}{m_b^2}-5\right)T^{\a\b}_a T^\g_{b\g}
\\\nonumber
&&+\left(\frac{m_b^2}{m_a^2}-5\right)T^{\a\b}_b T^\g_{a\g}
+4\left(T^{\a\g}_a T^\beta_{b\g}+T^{\a\g}_b T^\b_{a\g}\right)+(m_a^2+m_b^2)c^2\left(
N^\a_a N_b^\b+N^\b_aN^\a_b\right)
\\\label{38}
&&-2\left(T^{\a\b\g}_a N_{b\g}+T^{\a\b\g}_b N_{a\g}\right)-2\eta^{\a\b}\left[T^{\g\delta}_a T_{b\g\delta}-\left(\frac{m_b}{2m_a}+\frac{m_a}{2m_b}\right)^2T^{\g}_{a\g} T^\delta_{b\delta}\right]\Bigg\},
\een
where we have introduced the following abbreviation for the integral
\be
\mathcal{B}_n=\int_0^\pi \mathcal{F}(\Theta)\left(1-\cos^n\Theta\right)\sin\Theta d\Theta.
\ee{35b}

We note that with this methodology we were able to write the production terms (\ref{35c}) and (\ref{38}) in terms on the moments of the distribution function $N^{\a}_a$, $T^{\a\b}_a$ and $T^{\a\b\g}_a$, without the knowledge of the distribution function.

\section{Linearized 13$r$+1 field equations}

Once the constitutive equations for $T_a^{\a\b\g}$, $\pp_a^{\a}$ and  $\pp_a^{\a\b}$ are given  in terms of the 13$r$+1 fields (\ref{fi}) through (\ref{26}), (\ref{35c}) and (\ref{38}), respectively, we can obtain from the balance equations (\ref{10})
the corresponding 13$r$+1 field equations.
Here we are interested in the field equations that are linear in the non-equilibrium quantities  $\J_a^\alpha, \varpi_a, \q_a^\alpha, \p_a^{\langle \alpha\beta\rangle}$ and their derivatives. Hence,  the linearized  field equations for the partial fields particle number density $\n_a$, four-velocity $U^\alpha$ and temperature $T$ read
\ben\lb{46}
&&\D \n_a+\n_a\nabla_\a U^\a+\nabla_\a \J_a^\a=0,
\\\lb{46a}
&&\frac{\n\h}{c^2}\D U^\a+\nabla_\b \p^\lab-\nabla^\a\left(\p+\varpi\right)+\frac{1}{c^2}\D \q^\a=0,\\\lb{46b}
&&\n c_v\D\,T+\p\nabla^\a U_\a+\nabla^\a \q_\a=0.
\een
The balance equation of the particle number density  (\ref{46}) follows from (\ref{10}a) by using  (\ref{42}) of Appendix B. Note that we have introduced  the operators $\D$ and $\nabla_\a$ defined through the relationships
\be
\D=U^\a\partial_\a,\qquad \nabla_\a=\Delta_\a^\b\partial_\b,\qquad\hbox{so that}\qquad \partial_\a =\frac{1}{c^2}U_\a\D+\nabla_\a\qquad\hbox{and}\qquad U^\a\nabla_\a=0.
\ee{41}
For the  mixture the balance equations of the four-velocity (\ref{46a}) and of the temperature (\ref{46b}) follow from (\ref{10}b), (\ref{43}) and (\ref{39}) of the Appendix B and by summing the resulting equation over all constituents. The former refers to the projection of this equation by $\Delta^{\a\b}$, while the later is the projection with respect to $U^\a$. In (\ref{46b}) it was introduced  the heat capacity per particle at constant volume of the mixture
\be
c_v=\left(\frac{\partial e}{\partial T}\right)_v=\sum_{a=1}^rk\frac{\n_a}{\n}\left[\z_a^2+5G_a\z_a-G_a^2\z_a^2-1\right].
\ee{46c}

We get the balance equations for the partial diffusion fluxes $\J_a^\a$  from  (\ref{10}b), (\ref{43}) and (\ref{39}) of the Appendix B  by taking the projection $\Delta^{\a}_\b$ of the resulting equation and by subtracting the $r$th equation from the $a$th equation. This is necessary  in order to obtain $3(r-1)$ independent scalar balance equations. Hence it follows
\ben\no
&&\D \left(\frac{\J_a^\a}{\n_a}\right)-\D\left(\frac{\J_r^\a}{\n_r}\right)+\frac{c^2}{\n_a\h_a}\left[\nabla_\b \p_a^\lab-\nabla^\a\left(\p_a+\varpi_a\right)+\frac{1}{c^2}\D \q_a^\a\right]
-\frac{c^2}{\n_r\h_r}\bigg[\nabla_\b \p_r^\lab+\frac{1}{c^2}\D \q_r^\a
\\\no
&&\quad-\nabla^\a\left(\p_r+\varpi_r\right)\bigg]=-\frac{\pi}{2\n_a\h_a}\mathcal{B}_1\sum_{b=1}^r\bigg\{
\left[\h_a+\h_b-3kT\right]\left(\n_b\J_a^\a-\n_a\J_b^\a\right)+\n_b \q_a^\a-\n_a \q_b^\a\bigg\}
\\\label{47}
&&\quad-\frac{\pi}{2\n_r\h_r}\mathcal{B}_1\sum_{b=1}^r\bigg\{
\left[\h_b+\h_r-3kT\right]\left(\n_r\J_b^\a-\n_b\J_r^\a\right)+\n_r \q_b^\a-\n_b \q_r^\a\bigg\}.
\een

The balance equations for the partial non-equilibrium pressures $\varpi_a$, heat fluxes $\q_a^\a$ and pressure deviator tensors $\p_a^\lab$ are obtained from (\ref{10}c), together with (\ref{44}) and (\ref{40}) of the Appendix B as follows: first the projection $U_\a U_\b$ lead to the balance equations for the partial non-equilibrium pressures $\varpi_a$:
\ben\no
&&\frac{C_{2a}}{2}\D\varpi_a+\frac{m_a^2+C_{1a}}{2}\D\n_a-\frac{\zeta_a}{2T}\n_a C_{1a}^\prime\D T-5\frac{C_{3a}}{c^2}\nabla_\g \q_a^\g+\frac{\n_a}{6}(m_a^2+5C_{1a})\nabla_\g U^\g
\\\no
&&\quad+\frac{C_{5a}+C_{6a}}{c^2}\nabla_\g \J_a^\g=-\frac{\pi}{4c^4}\mathcal{B}_1\sum_{b=1}^r \left[\n_b\left(c^4C_{2a}+6\e_b\right)\varpi_a-\n_a\left(c^4C_{2b}+6\e_a\right)\varpi_b\right]
\\\no
&&\quad-\frac{\pi}{8c^4}\mathcal{B}_2\sum_{b=1}^r \bigg\{\n_b\varpi_a\bigg[c^4C_{2a}-3\left(5-
\frac{m_b^2}{m_a^2}\right)\e_b+6\left(\e_b-3kT\right)\left(\frac{m_a}{2m_b}+\frac{m_b}{2m_a}\right)^2
\\\no
&&\quad+6kT\bigg]+\n_a\varpi_b\bigg[c^4C_{2b}-3\left(5-
\frac{m_a^2}{m_b^2}\right)\e_a+6\left(\e_a-3kT\right)\left(\frac{m_a}{2m_b}+\frac{m_b}{2m_a}\right)^2
\\\no
&&\quad+6kT\bigg]+\frac{3\n_a\n_b}{2}\bigg[\left(m_a^2-m_b^2\right)\left(\e_a\e_b-3(kT)^2\right)-3kT\left(m_a^2\e_a-m_b^2\e_b\right)
\\\lb{47a}
&&\quad  -kT\left(m_a^2\e_b-m_b^2\e_a\right)\bigg]\frac{m_b^2-m_a^2}{m_a^2m_b^2}\bigg\},
\een
where we have introduced $C_{1a}^\prime=dC_{1a}(\z_a)/d\z_a$.
Next, the balance equations for the partial heat fluxes $\q_a^\a$ follow from the projection $\Delta_\gamma^\a U_\b$:
\ben\no
&&5C_{3a}\D \q_a^\a-\frac{c^4}{6}\left[\left(m_a^2-C_{1a}\right)\nabla^\a\n_a+\frac{\zeta_a}{T}\n_a C_{1a}^\prime\nabla^\a T-C_{2a}\nabla^\a \varpi_a\right]-c^2C_{4a}\nabla_\b \p_a^\lab
\\\no
&&\quad
-\left(C_{5a}+C_{6a}\right)\D \J_a^\a-\frac{c^2}{6}\left(m_a^2\n_a+5\n_a C_{1a}\right)\D U^\a=-\frac{\pi}{2c^2}\mathcal{B}_1\sum_{b=1}^r\Bigg\{c^2\n_b\J_a^\a\bigg[C_{5a}+C_{6a}
\\\no
&&\quad-\frac{(m_b^2-C_{1b})c^2}{6}+\frac{\h_a}{c^2}(\e_b-3kT)-\frac{(m_a^2-m_b^2)c^2}{2}\bigg]-c^2\n_a\J_b^\a\bigg[C_{5b}+C_{6b}-\frac{(m_a^2-C_{1a})c^2}{6}
\\\no&&\quad+\frac{\h_b}{c^2}(\e_a-3kT)+\frac{(m_a^2-m_b^2)c^2}{2}\bigg]
+\n_b \big(5c^2C_{3a}-\e_b+3kT\big)\q_a^\a-\n_a\big(5c^2C_{3b} -\e_a
\\\no&&\quad
+3kT\big)\q_b^\a
\Bigg\} -\frac{\pi}{8c^2}\mathcal{B}_2\sum_{b=1}^r \Bigg\{\n_b\Bigg[10C_{3a}c^2+4(\e_b-kT)-\left(5-\frac{m_a^2}{m_b^2}\right)(\e_b-3kT)\Bigg]\q_a^\a
\\\no
&&\quad+\n_a\Bigg[10C_{3b}c^2+4(\e_a-kT)-\left(5-\frac{m_b^2}{m_a^2}\right)(\e_a-3kT)\Bigg]\q_b^\a+2c^2\n_b\bigg[C_{5a}+C_{6a}
\\\no
&&\quad-\frac{(m_b^2-C_{1b})c^2}{6}+\left(5-\frac{m_a^2}{m_b^2}\right)\frac{\e_b-3kT}{2}\frac{\h_a}{c^2}-2\frac{\h_a}{c^2}(\e_b-kT)
-\frac{(m_a^2+m_b^2)c^2}{2} \bigg]\J_a^\a
\\\no
&&\quad+2c^2\n_a\bigg[C_{5b}+C_{6b}-\frac{(m_a^2-C_{1a})c^2}{6}+\left(5-\frac{m_b^2}{m_a^2}\right)\frac{\e_a-3kT}{2}\frac{\h_b}{c^2}-2\frac{\h_b}{c^2}(\e_a-kT)
\\\lb{47b}
&&\quad-\frac{(m_a^2+m_b^2)c^2}{2} \bigg]\J_b^\a\Bigg\}.
\een
Finally, the projection $\Delta_\gamma^\a \Delta_\delta^\b-\Delta_{\gamma\delta}\Delta^{\a\b}/3$ imply the balance equations for the partial pressure deviator tensors $\p_a^\lab$:
\ben\no
&&
C_{4a}\D\p_a^\lab+2C_{3a} \nabla^{\langle\a}\q_a^{\b\rangle}+2C_{5a} \nabla^{\langle\a}\J_a^{\b\rangle}+\frac{c^2}{3}\n_a\left(m_a^2- C_{1a}\right) \nabla^{\langle\a}U^{\b\rangle}
\\\no
&&\quad=-\frac{\pi}{2c^2}\mathcal{B}_1\sum_{b=1}^r\bigg\{\n_b\big(c^2C_{4a}+\e_b-3kT\big)\p_a^\lab-\n_a\left(c^2C_{4b}+\e_a-3kT\right)\p_b^\lab\bigg\}
\\\no
&&\quad -\frac{\pi}{8c^2}\mathcal{B}_2\sum_{b=1}^r\bigg\{\n_b\bigg[2c^2C_{4a}+\left(5-\frac{m_a^2}{m_b^2}\right)
\left(\e_b-3kT\right)+8kT\bigg]\p_a^\lab
\\\lb{47c}
&&\quad+\n_a\bigg[2c^2C_{4b}+\left(5-\frac{m_b^2}{m_a^2}\right)\left(\e_a-3kT\right)
+8kT\bigg]\p_b^\lab\bigg\}.
\een

Hence, the system of equations (\ref{46}), (\ref{46a}), (\ref{46b}), (\ref{47}), (\ref{47a}), (\ref{47b}) and (\ref{47c}) compose 13$r$+1 linearized field equations for the fields
(\ref{fi}).

\section{Six-field theory for a binary mixture}

In this section we shall restrict ourselves to a binary mixture characterized by the six scalar fields of particle number densities $\n_1$, $\n_2$, four-velocity $U^\a$ and temperature $T$,
whose balance equations (\ref{46}) -- (\ref{46b}) are written as
\ben\lb{52a}
&& \D \n_1+\n_1\nabla_\a U^\a+\nabla_\a \J_1^\a=0,\qquad \D \n_2+\n_2\nabla_\a U^\a+\nabla_\a \J_2^\a=0,
\\\lb{52b}
&&
\frac{\n\h}{c^2}\D U^\a+\nabla_\b \p^\lab-\nabla^\a\left(\p+\varpi\right)+\frac{1}{c^2}\D \q^\a=0,\qquad
\n c_v\D\,T+\p\nabla^\a U_\a+\nabla^\a \q_\a=0.\qquad
\een

In this case the pressure deviator tensors $\p_1^\lab$, $\p_2^\lab$, the non-equilibrium pressures $\varpi_1$, $\varpi_2$, the heat fluxes $\q_1^\a$, $\q_2^\a$ and the diffusion flux $\J_1^\a=-\J_2^\a$ are no longer variables, but constitutive quantities. To determine these constitutive quantities we shall rely on the remaining 21 scalar equations (\ref{47}) -- (\ref{47c}) and a method akin to the Maxwellian iteration procedure, which is often used in kinetic theory of gases. In this method the equilibrium values of the constitutive quantities -- namely,  $\p_1^\lab=\p_2^\lab$=0,  $\varpi_1=\varpi_2=0$,  $\q_1^\a=\q_2^\a=0$ and  $\J_1^\a=-\J_2^\a=0$ -- are inserted on the left-hand sides of the remaining 21 scalar equations and the first iterated values are obtained from the production terms, i.e., from the right-hand sides of these equations.

Following the methodology described in the Appendix C we get from:

\noindent \textbf{(i)}  two equations that follow from (\ref{47c})
\ben\lb{53a}
\p_1^\lab=2\mu_1\nabla^{\langle\a}U^{\b\rangle},\qquad \p_2^\lab=2\mu_2\nabla^{\langle\a}U^{\b\rangle};
\een

\noindent \textbf{(ii)}  two equations that follow from (\ref{47a})
\ben\lb{53b}
\varpi_1=-\eta_1\nabla^\a U_\a,\qquad \varpi_2=-\eta_2\nabla^\a U_\a;
\een

\noindent \textbf{(iii)} one equation and  two equations that follow from (\ref{47}) and (\ref{47b}),  respectively,
\ben\lb{53c}
\J_1^\a=-\J_2^\a=\mathcal{D}_{12}\,\d_1^\a+\mathcal{D}_T\,\nabla^\a \mathcal{T},\qquad\q_1^\a=\lambda_1\,\nabla^\a \mathcal{T}+\mathcal{D}_1\,\d_1^\a,\qquad
\q_2^\a=\lambda_2\,\nabla^\a \mathcal{T}+\mathcal{D}_2\,\d_1^\a.\quad
\een

Equations (\ref{53a}) and (\ref{53b}) represent the constitutive equations of a relativistic Newtonian fluid -- also known as the Navier-Stokes law -- and the associated transport coefficients $\mu_1, \mu_2$ and $\eta_1, \eta_2$ are identified as the coefficients of shear and bulk viscosities, respectively.
In (\ref{53c}) we have introduced  the generalized diffusion forces
\ben\lb{55}
\d_a^\a=\frac{1}{kT}\left(\nabla^\a\p_a-\frac{\n_a\h_a}{\n\h}\nabla^\a\p\right),\qquad a=1,2,
\een
that are restricted by the constraint $\sum_{a=1}^2\d_a^\a=0$, so that $\d_2^\a=-\d_1^\a$, and the relativistic temperature gradient
\ben\lb{55a}
\nabla^\a \mathcal{T}=\left(\nabla^\a T-\frac{T}{\n\h}\nabla^\a \p\right),
\een
which in the non-relativistic limiting case reduces to the temperature gradient $\nabla^\a T$. Hence, we may identify (\ref{53c}a) as the generalized Fick's law and (\ref{53c}b,c) as the generalized Fourier's law. The corresponding transport coefficients in these equations are recognized as the coefficients of diffusion $\mathcal{D}_{12}$, thermal-diffusion $\mathcal{D}_T$, thermal conductivity $\lambda_1, \lambda_2$ and diffusion-thermal $\mathcal{D}_1, \mathcal{D}_2$.

Now we are ready to obtain the constitutive equations necessary to convert the system of equations (\ref{52a}) and  (\ref{52b}) into a system of field equations for the six fields of  partial particle number densities $\n_1$, $\n_2$, four-velocity $U^\a$ and temperature $T$. Indeed, by using the definitions of the pressure deviator tensor, non-equilibrium pressure and heat flux of the mixture given by (\ref{18}) and (\ref{18a}) we get
\ben\lb{57}
\p^\lab=2\mu\nabla^{\langle\a}U^{\b\rangle},\qquad
\varpi=-\eta\nabla^\a U_\a,\qquad
\q^\a=\lambda' \nabla^\a \mathcal{T}+\mathcal{D}\d_1^\a.
\een
Here the transport coefficients of shear viscosity $\mu$, bulk viscosity $\eta$ and thermal conductivity $\lambda'$ of the mixture and the diffusion-thermal coefficient $\mathcal D$ read
\ben
&&\mu=\mu_1+\mu_2,\qquad\eta=\eta_1+\eta_2,\qquad\lambda'=\lambda_1+\lambda_2+\left(\h_1-\h_2\right)\mathcal{D}_T,\\
&&\mathcal{D}=\mathcal{D}_1+\mathcal{D}_2+(\h_1-\h_2)\mathcal{D}_{12}.
\een
We call attention to the fact that the true thermal conductivity of the mixture is defined as the ratio of the heat flux and the temperature gradient when there is no diffusion , i.e., when $\J_1^\a=-\J_2^\a=0$. In this case the we get from (\ref{53c}) that
\ben
\d_1^\a=-\frac{\mathcal{D}_T}{\mathcal{D}_{12}}\nabla^\a \mathcal{T},\qquad\hbox{hence}\qquad \q^\a=\lambda \nabla^\a \mathcal{T},
\een
where the true thermal conductivity of the mixture is given by
\ben
\lambda=\lambda_1+\lambda_2-\frac{\mathcal{D}_T}{\mathcal{D}_{12}}\left(\mathcal{D}_1+\mathcal{D}_2\right).
\een

The constitutive relation for the diffusion flux (\ref{53c}a) together with the ones for the pressure deviator tensor, non-equilibrium pressure and heat flux (\ref{57}) imply into the desired field equations for the six fields $(\n_1, \n_2, U^\a, T)$,  when they are inserted into the system of equations  (\ref{52a}) and  (\ref{52b}).

The expressions for the transport coefficients, even in the order up to $\epsilon$, are too large to be given here. Bellow we present only their expressions in the non- and ultra-relativistic limiting cases.

In the non-relativistic limiting case the thermal energy of the gas $kT$ is smaller than the particle rest energy $m_1c^2$, so that $\z_1\gg1$ and we obtain
\ben\no
\mathcal{D}_{12}&=&\frac{kT}{\pi m_1\n\,\mathcal{B}_1}\bigg\{1-\frac{(\mathcal{B}_1+\mathcal{B}_2)\n+2\n_1\,\mathcal{B}_1}{2\n(2\mathcal{B}_1+\mathcal{B}_2)}\,\epsilon-\frac{4\mathcal{B}_1
+\mathcal{B}_2}{2(2\mathcal{B}_1+\mathcal{B}_2)}\bigg[ 1
\\\lb{d1}
&-&\frac{\mathcal{B}_1^2(26\n_1+6\n_2)+\mathcal{B}_2^2(\n_1+3\n_2)+\mathcal{B}_1\mathcal{B}_2(13\n_1+11\n_2)}{2\n(2\mathcal{B}_1+\mathcal{B}_2)
(4\mathcal{B}_1+\mathcal{B}_2)}\,\epsilon\,\bigg]\frac{1}{\z_1}\bigg\}+\mathcal{O}\left(\epsilon^2,\frac{1}{\z_1^2}\right),\qquad
\\\lb{d2}
\mathcal{D}_{T}&=&-\frac{5(\mathcal{B}_1+\mathcal{B}_2)k\n_1\n_2}{4\pi m_1\n^2\,\mathcal{B}_2(2\mathcal{B}_1+\mathcal{B}_2)}\bigg\{\frac{1}{\z_1}-\frac{5\mathcal{B}_1^2+\mathcal{B}_1\mathcal{B}_2-\mathcal{B}_2^2}{(\mathcal{B}_1+\mathcal{B}_2)(2\mathcal{B}_1+\mathcal{B}_2)}
\frac{1}{\z_1^2}\bigg\}\epsilon+\mathcal{O}\left(\epsilon^2,\frac{1}{\z_1^3}\right),
\\\lb{d4}
\mathcal{D}&=&-\frac{kT\h_1^0}{\pi m_1\n\,\mathcal{B}_1\mathcal{B}_2}\bigg\{\mathcal{B}_1+\mathcal{B}_2-\frac{\mathcal{B}_1(6\mathcal{B}_1+7\mathcal{B}_2)}{4\mathcal{B}_1+2\mathcal{B}_2}\frac{1}{\z_1}
\bigg\}\epsilon+\mathcal{O}\left(\epsilon^2,\frac{1}{\z_1^2}\right),
\\\lb{d7}
\mu&=&\frac{2k T}{3\pi \mathcal{B}_2}\bigg\{1+\frac{1}{\z_1}\left(1-\frac{\n_2}{\n}\epsilon\right)-\frac{7}{3\z_1^2}\left(1-2\frac{\n_2}{\n}\epsilon\right)\bigg\}
+\mathcal{O}\left(\epsilon^2,\frac{1}{\z_1^3}\right),
\\\lb{d8}
\eta&=&\frac{5k T}{6\pi \mathcal{B}_2}\frac{1}{\z_1^2}\bigg\{1-2\frac{\n_2}{\n}\epsilon-\frac{25}{2\z_1}\left(1-3\frac{\n_2}{\n}\epsilon\right)\bigg\}+\mathcal{O}\left(\epsilon^2,\frac{1}{\z_1^4}\right),
\\\lb{d3}
\lambda&=&\lambda'=\frac{5\,k^2T}{2\pi m_1\,\mathcal{B}_2}\bigg\{1-\frac{\n_2}{\n}\epsilon-\frac{39}{8\z_1^2}\left(1-3\frac{\n_2}{\n}\epsilon\right)
\bigg\}+\mathcal{O}\left(\epsilon^2,\frac{1}{\z_1^3}\right),
\een
In the ultra-relativistic limiting case the thermal energy of the gas $kT$ is larger than the particle rest energy $m_1c^2$, so that $\z_1\ll1$ and we have
\ben\no
\mathcal{D}_{12}&=&\frac{6c^2(2\mathcal{B}_1+\mathcal{B}_2)}{\pi \n\,\mathcal{B}_1(26\mathcal{B}_1+11\mathcal{B}_2)}\bigg\{1-\frac{2(11\mathcal{B}_1
+5\mathcal{B}_2)}{3(26\mathcal{B}_1+11\mathcal{B}_2)}\bigg[ 1\\\lb{e1}
&+&\frac{4\mathcal{B}_1^2(15\n_1+7\n_2)+\mathcal{B}_1\mathcal{B}_2(55\n_1+29\n_2)+\mathcal{B}_2^2(13\n_1+7\n_2)}{2\n(11\mathcal{B}_1+5\mathcal{B}_2)
(2\mathcal{B}_1+\mathcal{B}_2)}\,\epsilon\,\bigg]{\z_1^2}\bigg\}+\mathcal{O}\left(\epsilon^2,\z_1^3\right),\qquad
\\\lb{e2}
\mathcal{D}_{T}&=&-\frac{c^2(6\mathcal{B}_1+11\mathcal{B}_2)\n_1\n_2}{3\pi \n^2\,T\mathcal{B}_2(26\mathcal{B}_1+11\mathcal{B}_2)}\epsilon\,{\z_1^2}+\mathcal{O}\left(\epsilon^2,\z_1^3\right),
\\\no
\mathcal{D}&=&-\frac{k T c^2}{\pi \n\,\mathcal{B}_2}\frac{26\mathcal{B}_1^2+55\mathcal{B}_1\mathcal{B}_2+26\mathcal{B}_2^2}{78\mathcal{B}_1+33\mathcal{B}_2}\z_1^2\epsilon
+\mathcal{O}\left(\epsilon^2,\z_1^3\right),
\\\no
\\\lb{e7}
\mu&=&\frac{4k T}{5\pi\mathcal{B}_2}\bigg\{1+\frac{\z_1^4}{80}\left[\frac{1}{4}+2\frac{\n_2}{\n}\epsilon+\left(\gamma+\ln\frac{\z_1}{2}\right)\left(1+4\frac{\n_2}{\n}\epsilon\right)\right]
+\mathcal{O}\left(\epsilon^2,\z_1^5\right),
\\\lb{e8}
\eta&=&\frac{k T\n_1}{108\pi\n\mathcal{B}_2}\z_1^4\bigg\{1+\frac{10\n_2(22\mathcal{B}_1+29\mathcal{B}_2)}{\n(2\mathcal{B}_1+\mathcal{B}_2)}
\epsilon\bigg\}+\mathcal{O}\left(\epsilon^2,\z_1^5\right),
\\\lb{e3}
\lambda&=&\lambda'=\frac{4c^2k}{3\pi \,\mathcal{B}_2}\bigg\{1-\frac{\z_1^2}{3}\left(1+2\frac{\n_2}{\n}\epsilon\right)\bigg\}
+\mathcal{O}\left(\epsilon^2,\z_1^3\right),
\een
From the expressions of the  transport coefficients (\ref{d1}) -- (\ref{e3}) we note that:
\begin{enumerate}
  \item they are valid up to the $\epsilon$-order, which was the approximation used when the masses of the particles of the constituents  are not too disparate, so that we have written $m_2=m_1(1+\epsilon)$ with $\epsilon$ being a small quantity;
  \item in all expressions relativistic corrections are given in terms of the parameter $\z_1$;
  \item the coefficients of thermal-diffusion $\mathcal{D}_T$ and diffusion-thermal $\mathcal{D}$ are of order $\epsilon$;
  \item the thermal-diffusion coefficient $\mathcal{D}_T$ given by (\ref{d2}) is of relativistic order, i.e., it vanishes in the non-relativistic limiting case. This is well known result in the literature that the thermal-diffusion coefficient vanishes for a mixture of Maxwellian particles (see e.g. \cite{CC,BE});
  \item the diffusion-thermal coefficient $\mathcal{D}$ is also of relativistic order, since its expression (\ref{d4}) depends on the rest energy $\h_1^0=m_1c^2$;
  \item within the $\epsilon-$order the coefficients $\lambda$ and $\lambda'$, associated with the thermal conductivity of the mixture, coincide;
  \item for a single component -- i.e., when $\x_2=0$ -- the transport coefficients of shear and bulk viscosities and thermal conductivity in the non-relativistic and ultra-relativistic limiting cases reduce to the ones given in \cite{CK,GK}.
\end{enumerate}

 In the case of the particles have the same rest masses $m_1=m_2$ the coefficients of thermal-diffusion $\mathcal{D}_T$ and diffusion-thermal $\mathcal{D}$ vanish and the diffusion coefficient becomes the self-diffusion coefficient $\mathcal{D}_{11}$  whose expression is given by
\ben
&&\mathcal{D}_{11}=\frac{-2c^2(2\mathcal{B}_1+\mathcal{B}_2)(10G_1+\z_1-G_1^2\z_1)}{\Delta},\qquad \hbox{where}
\\\no
&&\Delta=
\pi\n\mathcal{B}_1\big\{2\mathcal{B}_1\big[G_1^3\z_1^2(G_1\z_1-8)
+G_1(30+8\z_1^2)+\z_1(3+\z_1^2)+2G_1^2\z_1(1-\z_1^2)\big]\\
&&\quad+\mathcal{B}_2(30G_1+3\z_1-13G_1^2\z_1)\big\}.
\een
When $m_1=m_2$   the character of the mixture is exclusively owing to the difference of the particle number densities of the components. Furthermore the coefficients of thermal conductivity, shear and bulk viscosities of the mixture reduce to
\ben
\lambda=\frac{2kc^2\z(\z+5G-G^2\z)^2}{\pi\mathcal{B}_2(\z+10G-\z G^2)},\qquad
\mu=\frac{2kTG^2\z}{\pi\mathcal{B}_2(2G+\z+2G^2\z)},\\
\eta=\frac{kT\z(20G+3\z-13G^2\z-2G\z^2+2G^3\z^2)^2}{3\pi\mathcal{B}_2(1-5G\z-\z^2+G^2\z^2)^2(\z+5G-\z G^2)},
\een
which are the same as those of a single gas \cite{GK,CK}. The constitutive equations in the case where the masses of the particles are identical read
\ben
\J_1^\a=-\J_2^\a=\mathcal{D}_{11}\,\d_1^\a,\qquad \p^\lab=2\mu\nabla^{\langle\a}U^{\b\rangle},\qquad
\varpi=-\eta\nabla^\a U_\a,\qquad
\q^\a=\lambda \nabla^\a \mathcal{T}.
\een

In the next section we shall investigate the solutions of the six-field theory concerning the propagation of forced and free waves in a relativistic binary mixture and for that end we shall need the values of the transport coefficients: diffusion $\mathcal{D}_{12}$ thermal-diffusion $\mathcal{D}_{T}$, thermal conductivity $\lambda'$, diffusion-thermal $\mathcal{D}$, shear viscosity $\mu$ and bulk viscosity $\eta$.

\begin{figure}\vskip1cm
\begin{center}
\includegraphics[width=7cm]{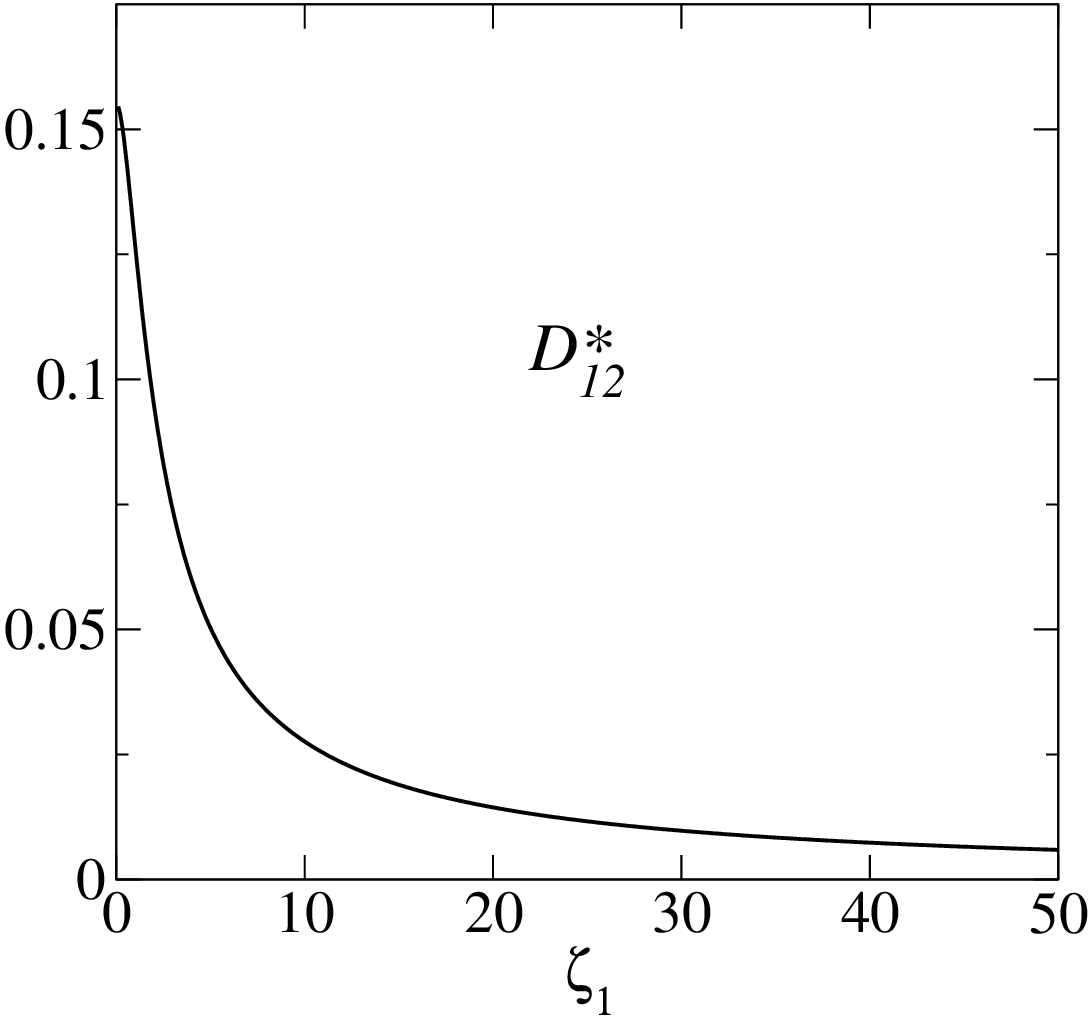}\hskip0.5cm
\includegraphics[width=7cm]{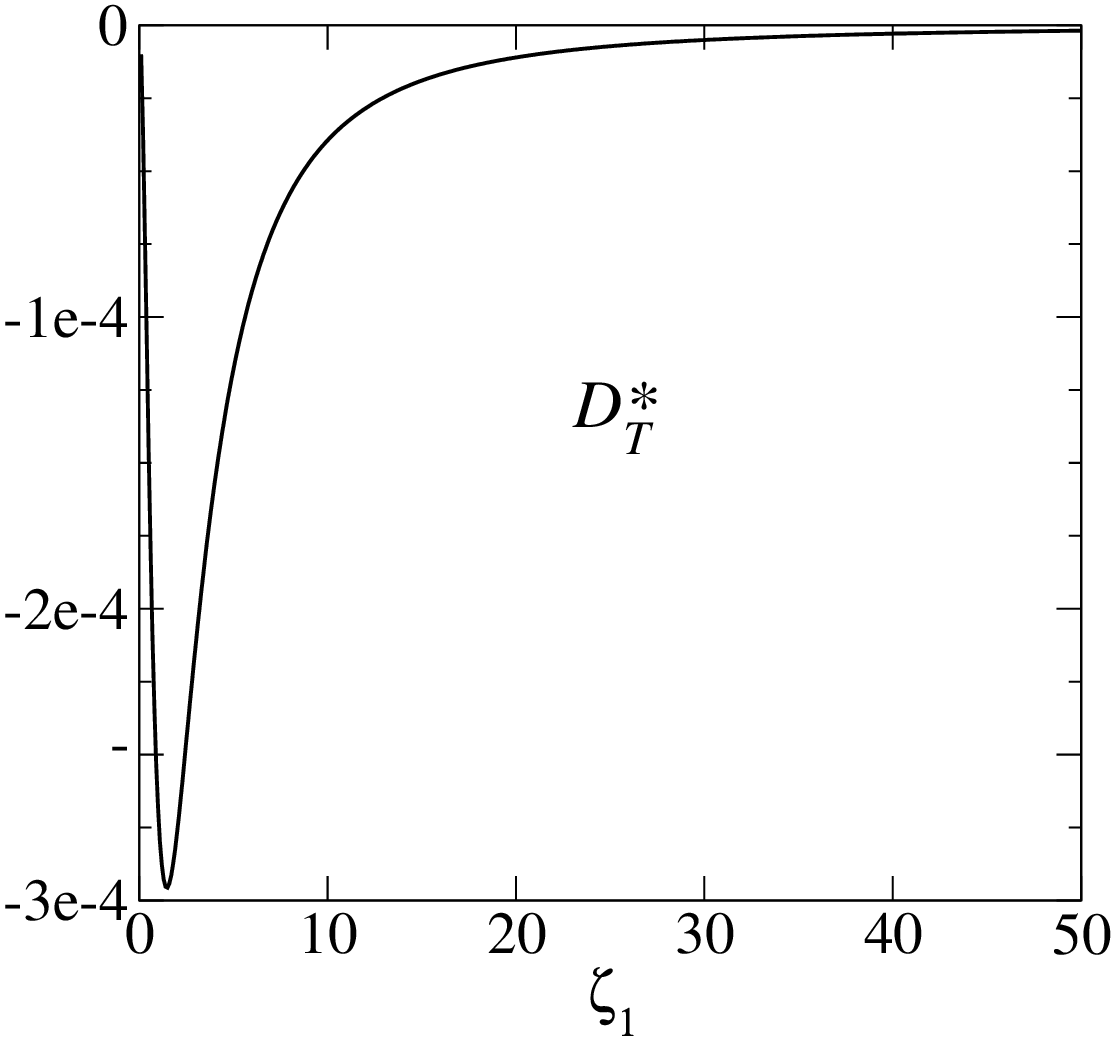}
\caption{Dimensionless coefficients as functions of $\z_1$: diffusion $\mathcal{D}_{12}^\ast$ (fig. 1a) and thermal-diffusion $\mathcal{D}_{T}^\ast$ (fig. 1b).}
\end{center}
\end{figure}

\begin{figure}\vskip1cm
\begin{center}
\includegraphics[width=7cm]{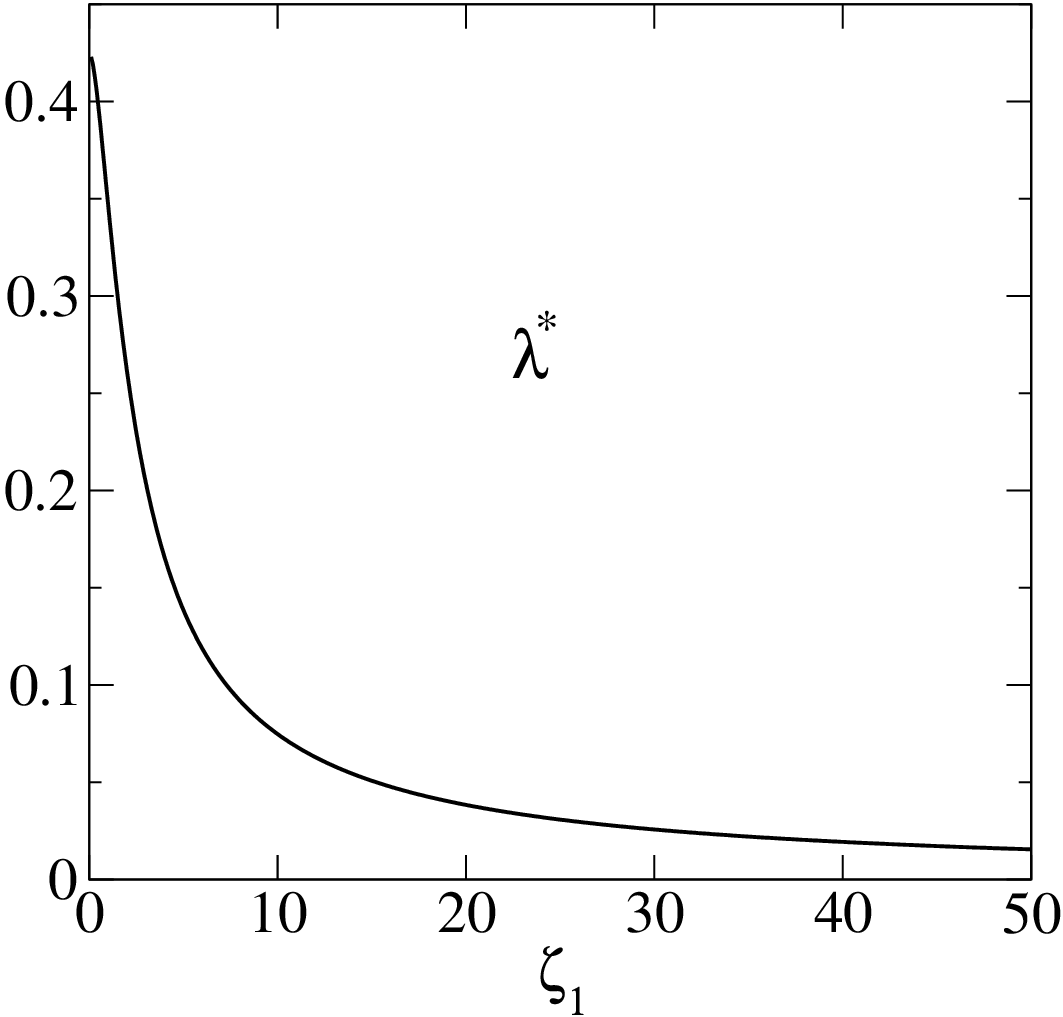}\hskip0.5cm
\includegraphics[width=7cm]{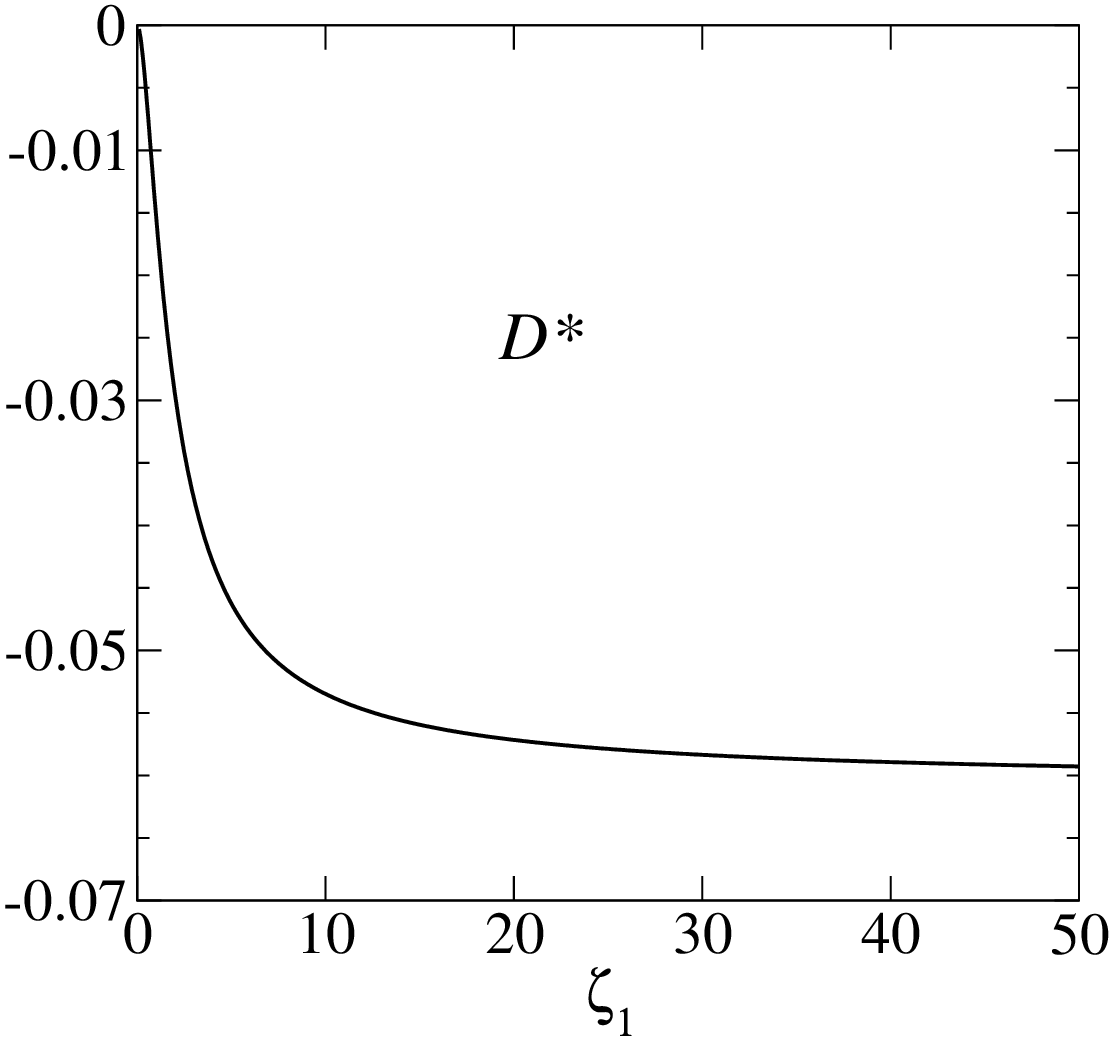}
\caption{Dimensionless coefficients as functions of $\z_1$: thermal conductivity $\lambda^\ast$ (fig. 2a) and diffusion-thermal $\mathcal{D}^\ast$ (fig. 2b).}
\end{center}
\end{figure}

\begin{figure}\vskip1cm
\begin{center}
\includegraphics[width=7cm]{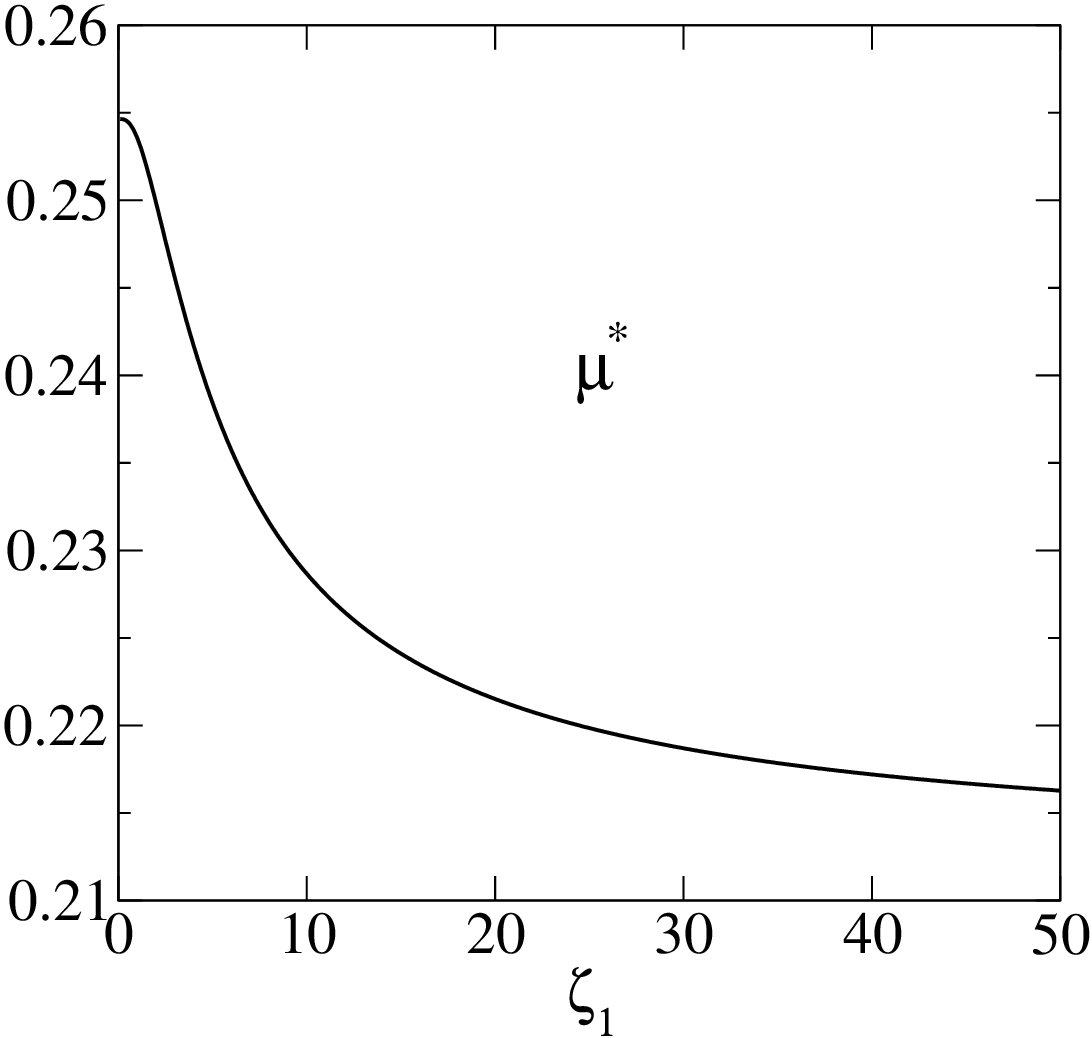}\hskip0.5cm
\includegraphics[width=7cm]{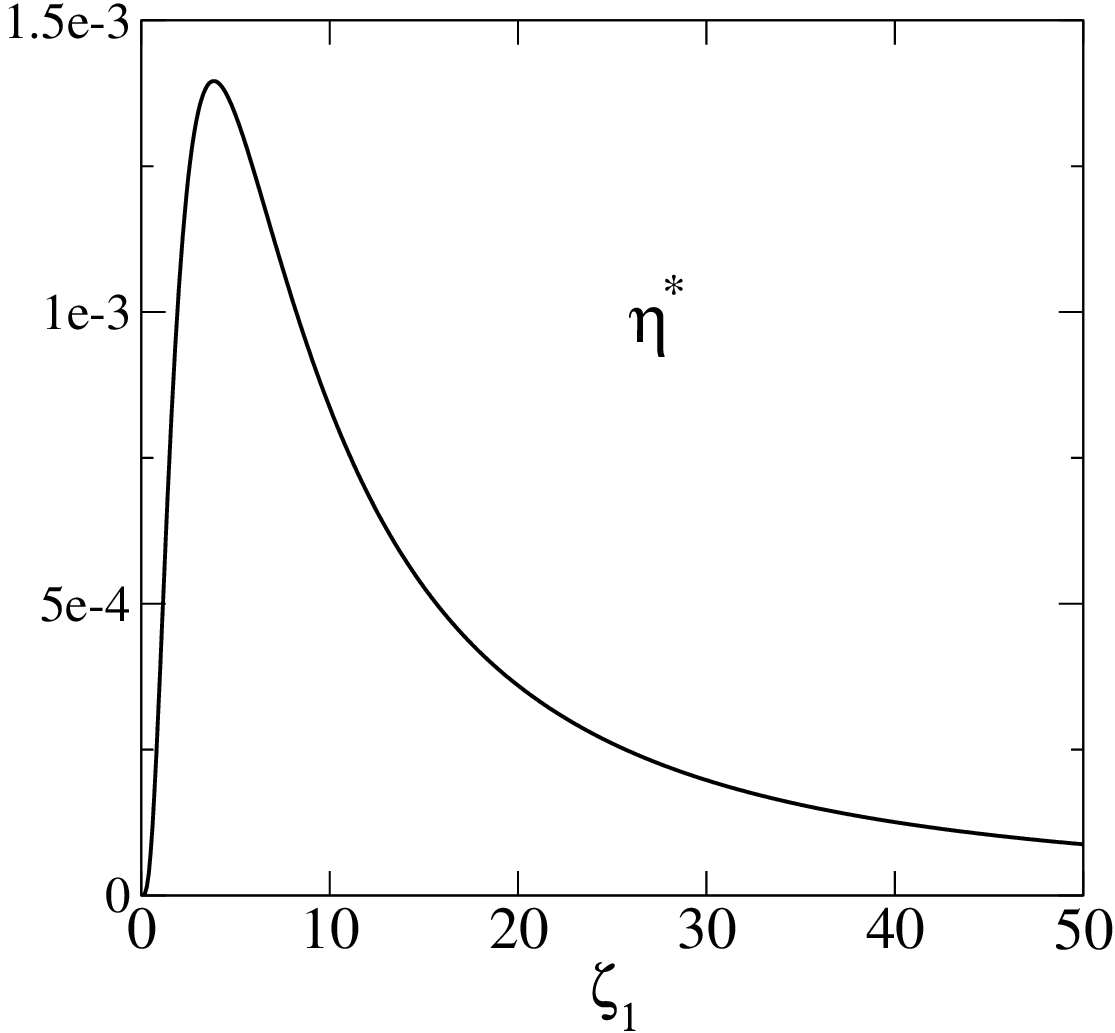}
\caption{Dimensionless coefficients as functions of $\z_1$: shear viscosity $\mu^\ast$ (fig. 3a) and bulk viscosity $\eta^\ast$ (fig. 3b).}
\end{center}
\end{figure}

In Figures 1 -- 3 are plotted  the dimensionless coefficients
\ben
\mathcal{D}_{12}^\ast=\mathcal{D}_{12}\frac{\mathcal{B}_2\n}{c^2},\qquad\mathcal{D}_{T}^\ast=\mathcal{D}_T\frac{\mathcal{B}_2T}{c^2},
\qquad\lambda^\ast=\lambda'\frac{\mathcal{B}_2}{k c^2},
\\\mathcal{D}^\ast=\mathcal{D}\frac{\mathcal{B}_2\n}{k T c^2},\qquad\mu^\ast=\mu\frac{\mathcal{B}_2}{k T },
\qquad\eta^\ast=\eta\frac{\mathcal{B}_2}{k T},
\een
as functions of the parameter $\z_1=m_1c^2/kT$. In these figures it was consider that the concentration of the constituent labeled with the index 1 was 60\%, i.e., $\x_1=\n_1/\n=0.6$ with $\x_2=1-\x_1$, and that the mass of the constituent 2 was 10\% larger than that of the constituent 1, so that $\epsilon=0.1$. Furthermore, the ratio of the integrals $\mathcal{B}_2/\mathcal{B}_1$ was taken equal to the ratio of the integrals that appear in the theory of non-relativistic Maxwellian particles, namely, $A_2(5)/A_1(5)$ (see, e.g. \cite{CC,BE}). Hence, we have adopted the value
$\mathcal{B}_2/\mathcal{B}_1=0.436/0.422$.

We infer from these figures for the dimensionless coefficients that:
\begin{enumerate}
\item the values of the diffusion $\mathcal{D}_{12}^\ast$, shear viscosity $\mu^\ast$ and thermal conductivity $\lambda^\ast$ in the non-relativistic limiting case are smaller than those in the ultra-relativistic one;
\item the coefficients of thermal-diffusion $\mathcal{D}_T^\ast$ and bulk viscosity $\eta^\ast$ vanish for very small and very large values of $\z_1$;
\item the  diffusion-thermal coefficient  $\mathcal{D}^\ast$  is very small in the ultra-relativistic limit and tends to a constant value in the non-relativistic limit.
\end{enumerate}

\section{Forced and free waves in a relativistic binary mixture}

As was pointed out in the last section,, the insertion of the constitutive relations (\ref{53c}a) and (\ref{57}) into the balance equations (\ref{52a}) and  (\ref{52b})
leads to system of field equations for the determination of the six fields of partial particle number densities $\n_a$ $(a=1,2)$,
mixture four-velocity $U^\alpha$ and mixture temperature $T$. Let us then look for solutions of form
\ben
\label{pert1}
\n_{a}=\x_{a} \n_{0}+{\overline {\n}}_{a}\exp\left(\imath q x-\imath\omega t\right),
\qquad
T=T_{ 0}+{\overline T}\exp\left(\imath q x-\imath \omega t\right),
\\
\label{pert3}
U^{\alpha}=\left(\frac{c}{\sqrt{1-\left[({\overline v}/c)
\exp\left(\imath q x-\imath\omega t\right)\right]^2}},
\frac{ {\overline v} \exp\left(\imath q x-\imath\omega t\right)}{\sqrt{ 1-\left[({\overline v}/c)
\exp\left(\imath q x-\imath\omega t\right)\right]^2}},0,0
\right),
\een
which represent small perturbations about an equilibrium state characterized by constant partial number densities $\x_a \n_0$, constant temperature
$T_0$ and vanishing value of the velocity in the longitudinal direction parallel to the $x$ axis. Besides, the amplitudes of the perturbed fields - i.e., the overlined quantities - are small
so that only linear deviations from equilibrium need to be take into account, while $\omega$ and $q$ are angular frequency and the wavenumber of the
perturbations, respectively.

In the linear case, we can easily verify that the operators $\D$ and  $\nabla^\alpha$ become $\D=\partial/\partial t$
and $\nabla^\alpha=(0,-\partial/\partial x^i)$. Hence, the insertion of expressions (\ref{pert1}) and (\ref{pert3}) into the field equations leads to
following system of algebraic equations for the amplitude of the perturbations
\be
\left(
  \begin{array}{cccc}
    M_{11}&M_{12}&M_{13}&M_{14} \\
    M_{21}&M_{22}&M_{23}&M_{24} \\
    M_{31}&M_{32}&M_{33}&M_{34} \\
    M_{41}&M_{42}&M_{43}&M_{44} \\
  \end{array}
\right)
\left(
  \begin{array}{c}
    \overline \n_1/\n_0 \\
    \overline \n_2/\n_0 \\
    \overline v/c \\
    \overline T/T_0 \\
  \end{array}
\right)=
\left(
  \begin{array}{c}
    0 \\
    0 \\
    0 \\
    0 \\
  \end{array}
\right),
\ee{w3}
where the elements of the matrix are given by
\ben\lb{w4}
&&M_{11}=\omega_\ast+\imath\left[\x_2\left(1+\epsilon\,\a_1\x_1\right)\mathcal{D}_{12}^\ast-\frac{\mathcal{D}_T^\ast}{\h_\ast}\right] q_\ast^2,\\
&&M_{12}=-\imath\left[\x_1 \left(1-\epsilon\,\a_1\x_2\right)\mathcal{D}_{12}^\ast+\frac{\mathcal{D}_T^\ast}{\h_\ast}\right] q_\ast^2,
\\
&& M_{13}=-\x_1 q_\ast,\qquad M_{14}=\imath\left[\epsilon\,\a_1\x_1\x_2\mathcal{D}_{12}^\ast +\frac{\mathcal{D}_T^\ast}{\h_\ast}(\h_\ast-1)\right] q_\ast^2,
\\
&&M_{21}=-\imath\left[\x_2\left(1+\epsilon\,\a_1\x_1\right)\mathcal{D}_{12}^\ast-\frac{\mathcal{D}_T^\ast}{\h_\ast}\right] q_\ast^2,
\\ && M_{22}=\omega_\ast+\imath\left[\x_1\left(1-\epsilon\,\a_1\x_2\right)\mathcal{D}_{12}^\ast +\frac{\mathcal{D}_T^\ast}{\h_\ast}\right] q_\ast^2,
\qquad M_{23}=-\x_2 q_\ast,
\\
&&  M_{24}=-M_{14},\qquad
 M_{31}=-q_\ast+\imath\left[\frac{\lambda^\ast}{\h_\ast}-\x_2\left(1+\epsilon\,\a_1\x_1\right)\mathcal{D}^\ast\right] q_\ast\omega_\ast,
\\
&& M_{32}=-q_\ast+\imath\left[\frac{\lambda^\ast}{\h_\ast}+\x_1\left(1-\epsilon\,\a_1\x_2\right)\mathcal{D}^\ast\right] q_\ast\omega_\ast,
\\
&&M_{33}=\h_\ast\omega_\ast+\imath \left(\eta^\ast+\frac{4}{3}\mu^\ast\right)q_\ast^2,
\een
\ben
&&M_{34}=-q_\ast-\imath\left[\frac{\lambda^\ast}{\h_\ast}(\h_\ast-1)+\epsilon\,\a_1\x_1\x_2\mathcal{D}^\ast\right] q_\ast\omega_\ast,\\
&&M_{41}=\imath\left[\x_2\left(1+\epsilon\,\a_1\x_1\right)\mathcal{D}^\ast -\frac{\lambda^\ast}{\h_\ast}\right] q_\ast^2,
\\\lb{w6}
&&M_{42}=-\imath\left[\x_1\left(1-\epsilon\,\a_1\x_2\right)\mathcal{D}^\ast +\frac{\lambda^\ast}{\h_\ast}\right] q_\ast^2,\qquad M_{43}=-q_\ast,
\\
&& M_{44}=c_v^\ast\,\omega_\ast+\imath\left[\frac{\lambda^\ast}{\h_\ast}(\h_\ast-1)
+ \epsilon\,\a_1\x_1\x_2\mathcal{D}^\ast\right]q_\ast^2.\qquad
\een
In the above equations we have introduced  a mean free time $\tau$ and starred quantities which correspond to dimensionless coefficients. They are given by
\ben\label{w7}
&&\tau=\frac{1}{n_0\mathcal{B}_2},\qquad q_\ast=q\,c\tau,\qquad \omega_\ast=\omega\tau,\qquad\\
&&  \h_\ast=\frac{\h}{kT}=\z_1G_1\left(1+\epsilon\,\a_1\x_2\right),\qquad\a_1=\frac{G_1^2\z_1-\z_1-4G_1}{G_1},
\een
\be
 c_v^\ast=\frac{c_v}{k}=\z_1^2+5\z_1G_1-\z_1^2G_1^2-1+\epsilon\,\x_2\z_1\left[2\z_1^2G_1\left(1-G_1^2\right)+\z_1\left(13G_1^2-3\right)-20G_1\right].
\ee{cv}

The system of algebraic equation (\ref{w3}) has a non-trivial solution if the determinant of the matrix of the coefficient vanishes. This condition leads to
a dispersion relation which can be used to study the propagation of sound waves, as well as the dynamical behaviour of small disturbances induced
by spontaneous internal fluctuations. For sound waves, the dispersion relation is solved by taking the angular frequency as a real input variable.
As a result a wavenumber is found which in general is complex. In this case, the phase velocity $v_p$ and the attenuation coefficient $\alpha$ are defined by
\begin{equation}
v_p=\frac{\omega}{\Re(q)}\quad\quad\text{and}\quad\quad \alpha=\Im(q).
\end{equation}
On the other hand, for the eigenmodes, we consider the wavenumber as real input and solve the dispersion relation to obtain
the complex angular frequency. The real part of the angular frequency $\Re(\omega)$ gives the oscillation frequency of a small internal perturbation with
wavenumber $q$, while its imaginary part $\Im(\omega)$ describes the decay in time of the oscillation amplitude.

\subsection{Acoustic Solution in the Low-Frequency Limit}

In the low-frequency limit, the acoustic solution of the dispersion relation can be determined by expanding the reduced wavenumber $q_\ast$ in power
series of the reduced oscillation frequency $\omega_\ast$ as
\begin{equation}
\label{exps}
q_\ast=a_0\,\omega_\ast+a_1\,\omega_\ast^2+\dots,
\end{equation}
where $a_0$, $a_1$, $\dots$ are complex expansion coefficients. By inserting the expansion (\ref{exps}) into the dispersion relation and equating
equal powers of the reduced oscillation frequency $\omega_\ast$ we get
\ben
\label{expa0}
&&a_0=\pm\sqrt{\frac{\h_\ast}{\gamma}},
\\
\label{expa1}
&&a_1=\imath \frac{a_{0}}{2 \gamma}\left\{\frac{4}{3} \mu^\ast+\eta^\ast
+\frac{\left[(\gamma-1) a_{0}^2-1\right]^2}{a_{0}^2} \lambda^\ast
+\gamma \left[(\gamma-1) a_{0}^2-1\right] \mathcal{D}^\ast\alpha_1\, \x_1 \x_2\,\epsilon\right\},
\een
where $\gamma=(c_v^\ast+1)/c_v^\ast$ is the specific heat ratio.

The non-relativistic limit for the phase velocity and the attenuation coefficient in the low-frequency regime can be derived
from the above expressions when we assume that the thermal energy of the gas mixture is smaller than the particle rest energy, i.e., when
$\zeta_1 \gg 1$. Hence, we have
\ben
\label{vp1}
&&\frac{v_{p}}{v_{0}}=\pm \left\{1-\frac{7}{4\zeta_1}\left(1-\x_{2}\epsilon\right)
+\dots\right\},
\\
\label{at1}
&&\frac{\alpha v_{0}}{\omega}=\pm \frac{7}{15\pi} (\omega\tau)\left\{1
+\frac{15}{28\z_1}\left(1-\x_{2}\epsilon\right)
+\dots\right\},
\een
where $v_0$ denotes the adiabatic sound speed for a non-relativistic binary gas mixture:
\begin{equation}
v_0=\sqrt{\frac{5}{3}\,\frac{kT_0}{m_1}\left(1-\x_2\epsilon\right)}
\end{equation}
At this point, it is important to mention that in the derivation of the above expressions for the phase velocity
and the attenuation coefficient we have assumed that the masses of the particles of the constituents are not too disparate so that
$\epsilon \ll 1$.

  When $\zeta_1\rightarrow \infty$, we verify
from expressions (\ref{vp1}) and (\ref{at1}) that the phase velocity reduces to
the adiabatic sound speed for a non-relativistic binary gas mixture, while the reduced attenuation coefficient $\alpha v_0/\omega$ varies
linearly with the reduced frequency.

On the other hand, the ultra-relativistic limit for the phase velocity and the attenuation coefficient follows when $\zeta_1 \ll 1$, i.e., when
the thermal energy of the gas mixture is larger than the particle rest energy. In this case, we have
\ben\label{vp2}
&&\frac{v_{ p}}{c}=\pm \frac{1}{\sqrt{3}}\left\{1-\frac{(1+2 \x_{ 2}\epsilon)}{24} \zeta_{ 1}^2+\dots\right\}
\\
\label{at2}
&&\frac{\alpha c}{\omega}=\pm \frac{2\sqrt{3}}{5 \pi} \left(\omega\tau\right)\biggl\{1+\mathcal{O}(\zeta_{ 1}^4)\biggr\}.
\een
When $\zeta_1\rightarrow 0$ we note that the phase velocity is of order of the speed of light $v_p=\pm c/\sqrt{3}$, while the reduced attenuation coefficient, as
in the non-relativistic limiting case, also varies linearly with the reduced oscillation frequency.

If we consider $\x_2=0$ in (\ref{vp1}) -- (\ref{at2})  we get the phase velocities and the attenuation coefficients for a single relativistic gas with Maxwellian particles. The expressions for the phase velocities (\ref{vp1}) and (\ref{vp2}) when $\x_2=0$ are the same as those in the work \cite{CGK}. However, the attenuation coefficients (\ref{at1}) and (\ref{at2}) agree with those in \cite{CGK} only in the first term without the corrections in the parameter $\z_1$. This fact can be easily understood, since here we are considering Maxwellian particles while in \cite{CGK} a kind of  hard-sphere potential was used. Due to the fact that the attenuation coefficient depends on the transport coefficients, which are functions on the type of potential, the  corrections in the parameter $\zeta_1$ should not agree.

\subsection{Eigenmodes in the Small Wavenumber Limit}

By solving the dispersion relation to obtain the reduced oscillation frequency $\omega_\ast$ as a function of the reduced wavenumber
$q_\ast$ we verify the existence of four longitudinal eigenmodes. Two of these hydrodynamic eigenmodes are sound modes describing sound propagation in opposite
directions parallel to the wavenumber, while the other two eigenmodes are non-propagating modes that gives rise to purely diffusive effects. Expressions for these
eigenmodes can be derived in the small wavenumber limit by expanding $\omega_\ast$ in power series of $q_\ast$ as
\begin{equation}
\label{expem}
\omega_\ast=a_0+a_1 \,q_\ast+a_2\,q_\ast^2+\dots.
\end{equation}
By inserting expansion (\ref{expem}) into the dispersion relation and equating
equal powers of $q_\ast$ we obtain for the sound modes:
\ben
&&a_0=0,\quad\quad\quad a_1=\pm \sqrt{\frac{\gamma}{\h_\ast}},
\\\label{ca1}
&&a_{2} = -\imath \frac{a_{1}^2}{2 \gamma}\left\{\frac{4}{3} \mu^\ast+\eta^\ast
+\frac{[a_{1}^2-(\gamma-1)]^2}{a_{1}^2} \lambda^\ast
-\gamma \frac{[a_{1}^2-(\gamma-1)]}{a_{1}^2} \mathcal{D}^\ast\alpha_1\, \x_1 \x_2\,\epsilon\right\},
\een
 and for the diffusive modes:
\ben
&&a_{0} = 0,\quad\quad\quad a_1=0,
\\
\label{ca2}
&&a_{2} = -\frac{\imath}{2}\left\{ \mathcal{D}_{12}^\ast+\frac{(\gamma-1)}{\gamma} \lambda^\ast\pm
\left[\left(\mathcal{D}^\ast_{12}-\frac{(\gamma-1)}{\gamma} \lambda^\ast\right)^{2}+4 \frac{(\gamma-1)}{\gamma}
\mathcal{D}^\ast \mathcal{D}_{T}^\ast\right]^{ 1/2}\right\}.
\een

 As mentioned in Section VII the thermal-diffusion coefficient $\mathcal{D}_T$ and
the diffusion-thermal coefficient $\mathcal{D}$ are of order $\epsilon$ so that the products $\mathcal{D}^\ast\epsilon$ in (\ref{ca1}) and  $\mathcal{D}^\ast\mathcal{D}_T^\ast$ in (\ref{ca2}) can be neglected. In this case, the two sound modes can be written as
\ben\label{ca3}
\omega\tau=\pm \sqrt{\frac{\gamma}{\h_\ast}}(qc\tau) -\imath \frac{1}{2 \h_\ast}\left\{\frac{4}{3} \mu^\ast+\eta^\ast
+\frac{[\gamma-\h_\ast(\gamma-1)]^2}{\h_\ast\gamma} \lambda^\ast
\right\}(qc\tau)^2+\dots\,,
\een
and the two diffusive modes as
\be
\omega\tau=-\imath \frac{(\gamma-1)}{\gamma} \lambda^\ast (qc\tau)^2+\dots
\quad\quad\quad \hbox{and} \quad\quad\quad
\omega\tau=-\imath \mathcal{D}_{12}^\ast(qc\tau)^2+\dots\,.
\ee{ca4}

We may observe from (\ref{ca4})  a decoupling between entropy fluctuations and concentration fluctuations. A similar result is observed in a non-relavistic
binary gas mixture of Maxwellian particles \cite{BY}, where thermal-diffusion is zero and the non-propagating eigenmodes are related to thermal diffusivity
and mass diffusion processes, respectively.

In the non-relativistic limit, the two sound modes read
\be
\omega\tau=\pm\biggl\{1-\frac{7}{4\z_1}\left(1-\x_{2}\epsilon\right)+\dots\biggr\}(qv_{0}\tau)
-\imath \frac{7}{15 \pi}\biggl\{1-\frac{33}{7\z_1}\left(1-\x_{2}\epsilon\right)+\dots\biggr\}(qv_{0}\tau)^2+\dots,
\ee{ev1}
while the two diffusive modes become
\ben
\label{ea1}
&&\omega\tau=-\imath \frac{3}{5 \pi}\left\{1-\frac{3}{2\z_1}\left(1-\x_{2}\epsilon\right)+\dots \right\}(qv_{0}\tau)^2+\dots,
\\
&&\omega\tau=-\imath \frac{3}{5 \pi} \mathcal{B}\left\{1-\frac{1}{2} \frac{(3+\mathcal{B})}{(2+\mathcal{B})}\left(1-2 \x_{2}\right)\epsilon\right.\nonumber\\
&&\left.-\frac{1}{2} \frac{(4+\mathcal{B})}{(2+\mathcal{B})}
\left[1-\frac{26-36 \x_{2}+(13-14 \x_{2})\mathcal{B}+\mathcal{B}^2}{2 (2+\mathcal{B}) (4+\mathcal{B})} \epsilon\right]\frac{1}{ \zeta_{1}}+\dots \right\}(qv_{0}\tau)^2+\dots\,.
\een
Above we have introduced the abbreviation $\mathcal{B}=\mathcal{B}_2/\mathcal{B}_1$.

The two sound modes  in the ultra-relativistic limiting case  are given by
\ben\nonumber
\label{ev2}
\omega\tau&=&\pm\frac{1}{\sqrt{3}}\left\{1-\frac{\left(1+2 \x_{ 2}\epsilon\right)}{24} \zeta_{ 1}^{2}+\dots\right\}(qc\tau)
\\&-&\imath \frac{2}{15 \pi}\left\{1-\frac{\left(1+2 \x_{ 2}\epsilon\right)}{8} \zeta_{ 1}^{2}+\dots\right\}(qc\tau)^2+\dots\,,
\een
and the two diffusive modes read
\ben
\label{ea2}
\omega\tau&=&-\frac{\imath}{3 \pi}\left\{1-\frac{5}{24}\left(1+2 \x_{ 2}\epsilon\right)\zeta_{ 1}^{2}+\dots \right\}(qc\tau)^2+\dots
\\\nonumber
\omega\tau&=&-\imath \frac{6 \mathcal{B}}{\pi} \frac{(2+\mathcal{B})}{(26+11 \mathcal{B})}
\biggl\{1-\frac{2}{3} \frac{(11+5 \mathcal{B})}{(26+11 \mathcal{B})}
\biggr[1
\\
&+&\frac{(60-32 \x_{ 2}+(55-26 \x_{ 2})\mathcal{B}+(13-6 \x_{ 2}) \mathcal{B}^2)}{2 (2+\mathcal{B})(26+11 \mathcal{B})} \epsilon\biggr]
\zeta_{ 1}^2+\dots\biggr\}(qc\tau)^2+\dots
\een

Note that fluctuations associated with the sound modes propagate with  velocities $v=\Re(\omega)/{q}$ given by
\ben
v=\pm v_0\left\{
1-\frac{7}{4\z_1}\left(1-\x_{2}\epsilon\right)+\dots\right\}, \qquad
v=\pm\frac{c}{\sqrt{3}}\left\{
1-\frac{1}{24}\left(1+2 \x_{ 2}\epsilon\right) \zeta_{ 1}^{2}+\dots\right\},
\een
 in the non-relativistic and in
the ultra-relativistic limiting cases, respectively.

The expression (\ref{ca3}) for the sound modes  and the one for the diffusive mode (\ref{ca4}a) are the same as those obtained in the work \cite{KM} for a single relativistic gas. If we set $\x_2=0$ in (\ref{ca3}) and in (\ref{ca4}a)  we get the sound and diffusive modes of a single relativistic gas with Maxwellian particles. As was pointed out in the last subsection the relativistic corrections in the parameter $\z_1$ should not agree with those in \cite{KM}, since in the latter work a kind of hard-sphere potential was used.

The frequency and wavenumber regions of validity of the acoustic and eigenmodes solutions derived from the six field
theory for a binary mixture can be determined by imposing, respectively, the conditions (see \cite{KM}):
\ben
\frac{p}{\mu\omega} > 1,\qquad\hbox{and}\qquad \frac{q\mu c}{p} < 1,
\een
where $\mu$ is the shear viscosity of the mixture in the non-relativistic limiting case. The first condition tell us that
an acoustic solution based on the classical hydrodynamic description is valid as long as the molecular collision frequency is
larger than the wave frequency, while the second one tell us that the eigenmode solution is valid as long as the mean free path
of the molecules is smaller than the wavelength of the internal fluctuations. In terms of the dimensionless frequency
$\omega_\ast=\omega\tau$ and the dimensionless wavenumber $q_\ast=qc\tau$ we can rewrite the above conditions as
\ben
\omega_\ast > \frac{3\pi}{2},\qquad\hbox{and}\qquad q_\ast < \frac{3\pi}{2}.
\een
Finally, it is important to mention that another possible way to test the range of validity of an extended
hydrodynamic description for the acoustic problem in relativistic gases is to consider how the maximum speed of
propagation approaches the light speed with the increasing number of the moments. For the relativistic single gas case,
a complete analysis based on this method was given in the work \cite{Ru2}.

\section{Conclusions}

To sum up a mixture of $r$ constituents was analyzed within the framework of Boltzmann equation by using Grad's moment method. The mixture was described by $13r+1$ basic fields of  four-velocity, temperature of the mixture, particle number densities, diffusion fluxes, non-equilibrium pressures, heat fluxes and pressure deviator tensors. The $13r+1$ field equations were obtained from Grad's distribution function applied to a mixture of Maxwellian particles where the rest masses of the constituents are not too disparate. From the system of $13r+1$ field equations it was determined the constitutive equations for a  binary mixture with six  scalar fields of particle number densities, four-velocity and temperature. The generalized laws of Fick, Fourier and Navier-Stokes were obtained and the transport coefficients of diffusion, thermal-diffusion, diffusion-thermal, thermal conductivity and bulk and shear viscosities were determined. Explicit expressions for these coefficients were given in the non-relativistic and ultra-relativistic limiting cases as well as their graphs in terms of a parameter which represents the ratio of the rest energy of a particle and the thermal energy of the gas mixture. An analysis of forced and free waves for the binary mixture was also performed. The phase velocity and attenuation coefficient were determined for the acoustic solution in the low frequency limit. It was also shown that there exist two sound modes describing sound propagation in opposite directions parallel to the wavenumber, and two non-propagating modes related to purely diffusive effects.

\acknowledgments
This paper is dedicated to the memory of Dr. Leopoldo  Garc\'{\i}a-Col\'{\i}n Scherer.
The work of G.M.K. has been supported by the Conselho Nacional de Desenvolvimento Cient\'{\i}fico e
Tecnol\'ogico (Brazil).

\appendix

\section{Evaluation of the production terms}

Let us evaluate  the production term $\pp_a^\b$. By using the relationships (\ref{30a}) and the representation (\ref{34}), the production term (\ref{10a}) becomes
\ben\label{35a}\no
\pp_a^\b&=&\frac{c}{2}\sum_{b=1}^r\int (Q^{\prime\b}-Q^\b)f_a f_b F_{ba}\sigma_{ab} d\Omega\frac{d^3 p_b}{ p_{b0}}\frac{d^3 p_a}{p_{a0}}
\\
&=&\frac{c}{2}\sum_{b=1}^r\int Q(\k^{\prime\b}-\k^\b)f_a f_b F_{ba}\sigma_{ab} d\Omega\frac{d^3 p_b}{ p_{b0}}\frac{d^3 p_a}{p_{a0}}.
\een
If we write the element of solid angle as $d\Omega=\sin\Theta d\Theta d\Phi$ where $\Theta$ and $\Phi$ are the spherical angles of $\k^{\prime\b}$ with respect to $\k^\b$ and  use the following result
 \ben\label{A1a}
 \int_0^{2\pi}\int_0^\pi(\k^{\prime\b}-\k^\b)\sin\Theta d\Theta d\Phi=-2\pi \int_0^\pi \k^\b(1-\cos\Theta)\sin\Theta d\Theta,
 \een
 the production term (\ref{35a}) reduces to
 \ben\nonumber
\pp_a^\b=-\pi c\sum_{b=1}^r\int_0^\pi\int Q\,\k^\b f_a f_b F_{ba}\sigma_{ab} d\Omega\frac{d^3 p_b}{ p_{b0}}\frac{d^3 p_a}{p_{a0}}(1-\cos\Theta)\sin\Theta d\Theta
\\\label{35f}
=-\pi c\sum_{b=1}^r\int_0^\pi\int \bigg[Q^\b-\frac{(m_a^2-m_b^2)c^2}{P^2}P^\b\bigg] f_a f_b F_{ba}\sigma_{ab} d\Omega\frac{d^3 p_b}{ p_{b0}}\frac{d^3 p_a}{p_{a0}}(1-\cos\Theta)\sin\Theta d\Theta,\qquad
\een
thanks to (\ref{31a}).
Now if we take into account (\ref{34}) and transform the variables $(P^\a,Q^\a)$ into $(p_a^\a, p_b^\a)$ by using the relationships (\ref{28a}), we may  perform the integrations of (\ref{35a}) with respect to $\frac{d^3 p_b}{p_{b0}}\frac{d^3 p_a}{p_{a0}}$ and get
\be
\pp_a^\b=-\frac{\pi}{2c^2}\mathcal{B}_1\sum_{b=1}^r\left[\left(T^{\b\a}_a-T^\g_{a\g}\eta^{\b\a}\right)N_{b\a}
-\left(T^{\b\a}_b-T^\g_{b\g}\eta^{\b\a}\right)N_{a\a}\right].
\ee{35d}
In the above expression $\mathcal{B}_1$ is the following integral when $n=1$:
\be
\mathcal{B}_n=\int_0^\pi \mathcal{F}(\Theta)\left(1-\cos^n\Theta\right)\sin\Theta d\Theta.
\ee{35g}

The same methodology may be used to determine the production term $\pp_a^{\a\b}$. First we obtain from (\ref{10b})
\ben\label{36a}\no
\pp_a^{\a\b}=\frac{c}{4}\sum_{b=1}^r\int \big[(Q^{\prime\a}Q^{\prime\b}-Q^\a Q^\b)+P^\a (Q^{\prime\b}-Q^\b)
\\+P^\b (Q^{\prime\a}-Q^\a)\big]f_a f_b F_{ba}\sigma_{ab} d\Omega\frac{d^3 p_b}{ p_{b0}}\frac{d^3 p_a}{p_{a0}}.
\een
Next by using the relationship
 \ben\no
 \label{A1b}
&&\int_0^{2\pi}\int_0^\pi
\left(\k^{\prime\alpha}\k^{\prime\beta}
-\k^{\alpha}\k^{\beta}\right)\sin\Theta d\Theta d\Phi\\
&&=\pi \int_0^\pi \bigg(\frac{P^{\alpha}P^{\beta}}{P^2}
-\eta^{\alpha\beta} -3\,{\k^{\alpha}\k^{\beta}}\bigg)\left(1-\cos^2\Theta\right)\sin\Theta d\Theta,
\een
and (\ref{A1a}) it follows that the production term (\ref{36a}) reduces to
\ben\nonumber
&&\pp_a^{\a\b}=\frac{\pi c}{4}\sum_{b=1}^r\int_0^\pi \sin\Theta d\Theta\int\bigg\{\bigg[\frac{Q^2 P^\a P^\b}{P^2}-Q^2\eta^{\a\b}-3Q^\a Q^\b
+\frac{3(m_a^2-m_b^2)c^2}{P^2}(P^\a Q^\b
\\\no
&&\quad+P^\b Q^\a)\bigg](1-\cos^2\Theta)-2\bigg[P^\a Q^\b+P^\b Q^\a-\frac{(m_a^2-m_b^2)c^2}{P^2}\big[P^\b(P^\a-Q^\a)
\\\label{36b}
&&\quad+P^\a(P^\b-Q^\b)\big]\bigg](1-\cos\Theta)\bigg\}
f_a f_b F_{ba}\sigma_{ab} d\Omega\frac{d^3 p_b}{ p_{b0}}\frac{d^3 p_a}{p_{a0}}.
\een
By performing the integration with respect to $\frac{d^3 p_b}{p_{b0}}\frac{d^3 p_a}{p_{a0}}$ leads to
\ben\nonumber
&&\pp_a^{\a\b}=-\frac{\pi}{2c^2}\mathcal{B}_1\sum_{b=1}^r\Bigg\{T^{\a\b\g}_a N_{b\g}-T^{\a\b\g}_b N_{a\g}
-\frac{\left(m_a^2-m_b^2\right)c^2}{2}\left(N_a^\a N_b^\b+N_a^\b N_b^\a\right)+T^{\a\b}_a T^\g_{b\g}
\\\no
&&-T^{\a\b}_b T^\g_{a\g}\Bigg\}+\frac{\pi}{8c^2}\mathcal{B}_2\sum_{b=1}^r\Bigg\{\left(\frac{m_a^2}{m_b^2}-5\right)T^{\a\b}_a T^\g_{b\g}
+\left(\frac{m_b^2}{m_a^2}-5\right)T^{\a\b}_b T^\g_{a\g}
+4\big(T^{\a\g}_a T^\beta_{b\g}
\\\nonumber
&&+T^{\a\g}_b T^\b_{a\g}\big)-
2\left(T^{\a\b\g}_a N_{b\g}+T^{\a\b\g}_b N_{a\g}\right)+(m_a^2+m_b^2)c^2\left(
N^\a_a N_b^\b+N^\b_aN^\a_b\right)
\\\label{38a}
&&-2\eta^{\a\b}\left[T^{\g\delta}_a T_{b\g\delta}-\left(\frac{m_b}{2m_a}+\frac{m_a}{2m_b}\right)^2T^{\g}_{a\g} T^\delta_{b\delta}\right]\Bigg\}.
\een

It is worth to call attention that (\ref{A1a}) and (\ref{A1b}) result from the integration in the spherical angles, when the relative momentum four-vector after collision is expressed
in terms of the one before collision. These are general expressions which are valid for all representations of the distribution function, like those that appear in the Chapman-Enskog and Grad methods.

\section{Linearized  balance equations}

By  introducing the operators $\D$ and $\nabla_\a$ defined by the relationship
\be
\D=U^\a\partial_\a,\qquad \nabla_\a=\Delta^{\a\b}\partial_\b,\qquad\hbox{so that}\qquad \partial_\a =\frac{1}{c^2}U_\a\D+\nabla_\a,\qquad\hbox{and}\qquad U^\a\nabla_\a=0,
\ee{41a}
the linearized left-hand sides of (\ref{10}) read
\ben\lb{42}
&&\partial_\a N_a^\a=\D \n_a+\n_a\nabla_\a U^\a+\nabla_\a \J^\a_a,
\\\nonumber
&&\partial_\b T^{\a\b}_a=\frac{\n_a\h_a}{c^2}\D U^\a+\nabla_\b \p_a^\lab-\nabla^\a\left(\p_a+\varpi_a\right)+\frac{1}{c^2}\D \left(\q_a^\a+\h_a \J_a^\a\right)
\\\lb{43}&&\qquad
+\frac{U^\a}{c^2}\left[\D \left(\n_a\e_a\right)+\n_a\h_a\nabla_\b U^\b+\nabla_\b  \left(\q_a^\b+\h_a \J_a^\b\right)\right],
\\\nonumber
&&\partial_\g T^{\a\b\g}_a=U^\a U^\b\Bigg\{\frac{1}{3}\left[\left(m_a^2+2C_{1a}\right)\D\n_a-2\zeta_a\n_a C_{1a}^\prime\frac{\D T}{T}+2C_{2a}\D\varpi_a \right]+\n_a C_{1a}\nabla_\g U^\g
\\\nonumber
&&\quad-\frac{6C_{3a}}{c^2}\nabla_\g \q_a^\g+\frac{C_{6a}}{c^2}\nabla_\g \J_a^\g\Bigg\}+\frac{c^2}{6}\left[\left(m_a^2-C_{1a}\right)\D\n_a+\zeta_a\n_a C_{1a}^\prime\frac{\D T}{T}-C_{2a}\D\varpi_a\right]\eta^{\a\b}
\\\no
&&\quad+\frac{1}{6}\left(m_a^2\n_a+5\n_a C_{1a}\right)\left(U^\a\D U^\b+U^\b \D U^\a\right)+\frac{c^2}{6}\bigg[\left(m_a^2-C_{1a}\right)\nabla_\g\n_a+\zeta_a\n_a C_{1a}^\prime\frac{\nabla_\g T}{T}
\\\no
&&\quad-C_{2a}\nabla_\g\varpi_a\bigg]\left(\eta^{\a\g}U^\b+\eta^{\b\g}U^\a\right)
-\frac{5}{c^2}C_{3a}\left(U^\a\D \q_a^\b+U^\b\D \q_a^\a\right)+C_{3a}\big(\eta^{\a\b}\nabla_\g \q_a^\g+\nabla^\a \q^\b_a
\\\no
&&\quad+\nabla^\b q^\a_a\big)+\frac{c^2}{6}\big(m_a^2\n_a-\n_a C_{1a}\big)\big(\eta^{\a\b}\nabla_\g U^\g+\nabla^\a U^\b+\nabla^\b U^\a\big)+C_{4a}\big(\D \p_a^{\lab}+U^\b \nabla_\g \p_a^{\langle \a\g\rangle}
\\\lb{44}
&&\quad+U^\a \nabla_\g \p_a^{\langle \b\g\rangle}\big)+C_{5a}\left(\eta^{\a\b}\nabla_\g \J_a^\g+\nabla^\a \J^\b_a+\nabla^\b \J^\a_a\right)+\frac{C_{5a}+C_{6a}}{c^2}\left(U^\a\D \J_a^\b+U^\b\D \J_a^\a\right),\qquad
\een
thanks to the representations (\ref{13}), (\ref{16})  and (\ref{26}). Above we have introduced the notation $C_{1a}^\prime=dC_{1a}(\z_a)/d\z_a$.

Again by using  (\ref{13}), (\ref{16})  and (\ref{26}) we obtain from (\ref{35d}) and (\ref{38a}) the linearized production terms as functions of the 13$r$+1 moments of the distribution function:
\ben\lb{39}\no
&&\pp_a^\b=-\frac{\pi}{2c^2}\mathcal{B}_1\sum_{b=1}^r\Big\{\left[\n_b(\h_a+\h_b)-3\p_b\right]\J_a^\a-\left[\n_a(\h_a+\h_b)-3\p_a\right]\J_b^\a
\\
&&+\n_b \q_a^\a-\n_a \q_b^\a
+3\left(\n_b\varpi_a-\n_a\varpi_b\right)U^\a\Big\},
\een
$$
\pp_a^{\a\b}=-\frac{\pi}{2c^2}\mathcal{B}_1\sum_{b=1}^r\Bigg\{\frac{c^2}{3}\big[(\n_am_a^2+2\n_a C_{1a}+2C_{2a}\varpi_a)\n_b-(\n_b m_b^2+2\n_b C_{1b}$$
$$
+2C_{2b}\varpi_b)\n_a\big]U^\a U^\b+\frac{c^4}{6}\big[(\n_am_a^2-\n_a C_{1a}-C_{2a}\varpi_a)\n_b-(\n_b m_b^2-\n_b C_{1b}-C_{2b}\varpi_b)\n_a\big]\eta^{\a\b}$$
$$
-5\Big[C_{3a}\n_b(U^\a \q^\b_a+U^\b \q^\a_a)-C_{3b}\n_a(U^\a \q^\b_b+U^\b \q^\a_b)\Big]+c^2\Big[
C_{4a} \n_b \p^\lab_a-C_{4b} \n_a \p^\lab_b\Big]$$
$$
+\n_b\bigg[C_{5a}+ C_{6a}-\frac{c^2}{6}( m_b^2- C_{1b})\bigg]\big(U^\a \J_a^\b
+U^\b \J_a^\a\big)
-\n_a\bigg[C_{5b}+ C_{6b}-\frac{c^2}{6}(m_a^2- C_{1a})\bigg]\big(U^\a \J_b^\b$$
$$
+U^\b \J_b^\a\big)+(\n_b\e_b-3\p_b)\Big[\p_a^\lab-(\p_a+\varpi_a)\Delta^{\a\b}+\frac{1}{c^2}U^\b(\q_a^\a+\h_a \J_a^\a)+\frac{1}{c^2}U^\a(\q_a^\b+\h_a \J_a^\b)$$
$$
+\frac{\n_a\e_a}{c^2}U^\a U^\b\Big]-(\n_a\e_a-3\p_a)\Big[\p_b^\lab-(\p_b+\varpi_b)\Delta^{\a\b}+\frac{1}{c^2}U^\b(\q_b^\a+\h_b \J_b^\a)+\frac{1}{c^2}U^\a(\q_b^\b$$
$$
+\h_b \J_b^\b)+\frac{\n_b\e_b}{c^2}U^\a U^\b\Big]-3\varpi_b\bigg(\frac{\n_a\e_a}{c^2}U^\a U^\b-\p_a\Delta^{\a\b}\bigg)+3\varpi_a\left(\frac{\n_b\e_b}{c^2}U^\a U^\b-\p_b\Delta^{\a\b}\right)$$
$$
-\frac{(m_a^2-m_b^2)c^2}{2}\left[2\n_a\n_bU^\a U^\b+U^\a(\n_a\J_b^\b+\n_b\J_a^\b)+U^\b(\n_a\J_b^\a+\n_b\J_a^\a)\right]\Bigg\}$$
$$
-\frac{\pi}{8c^2}\mathcal{B}_2\sum_{b=1}^r\Bigg\{\frac{2c^2}{3}\left[(\n_am_a^2+2\n_a C_{1a}+2C_{2a}\varpi_a)\n_b+(\n_b m_b^2+2\n_b C_{1b}+2C_{2b}\varpi_b)\n_a\right]U^\a U^\b$$
$$
+2c^2\left[C_{4a} \n_b \p^\lab_a
+C_{4b} \n_a \p^\lab_b\right]
+\frac{c^4}{3}\big[(\n_am_a^2-\n_a C_{1a}-C_{2a}\varpi_a)\n_b+(\n_b m_b^2-\n_b C_{1b}$$
$$
-C_{2b}\varpi_b)\n_a\big]\eta^{\a\b}-10\big[C_{3a}\n_b(U^\a \q^\b_a+U^\b \q^\a_a)
+C_{3b}\n_a(U^\a \q^\b_b+U^\b \q^\a_b)\big]
+2\n_b\bigg[C_{5a}$$
$$
+ C_{6a}-\frac{c^2}{6}( m_b^2- C_{1b})\bigg]\Big(U^\a \J_a^\b+U^\b \J_a^\a\Big)
+2\n_a\Big[C_{5b}+ C_{6b}-\frac{c^2}{6}( m_a^2- C_{1a})\Big]\Big(U^\a \J_b^\b$$
$$
+U^\b \J_b^\a\Big)
+\left(5-\frac{m_a^2}{m_b^2}\right)\Big[(\n_b\e_b-3\p_b)\Big(\p_a^\lab-(\p_a+\varpi_a)\Delta^{\a\b}+\frac{1}{c^2}U^\b(\q_a^\a+\h_a \J_a^\a)$$
$$
+\frac{1}{c^2}U^\a(\q_a^\b+\h_a \J_a^\b)+\frac{\n_a\e_a}{c^2}U^\a U^\b\Big)-3\varpi_b\left(\frac{\n_a\e_a}{c^2}U^\a U^\b-\p_a\Delta^{\a\b}\right)\Big]+\left(5-\frac{m_b^2}{m_a^2}\right)\Big[(\n_a\e_a$$
$$
-3\p_a)\left(\p_b^\lab-(\p_b+\varpi_b)\Delta^{\a\b}+\frac{1}{c^2}U^\b(\q_b^\a+\h_b \J_b^\a)+\frac{1}{c^2}U^\a(\q_b^\b+\h_b \J_b^\b)+\frac{\n_b\e_b}{c^2}U^\a U^\b\right)$$
$$
-3\varpi_a\left(\frac{\n_b\e_b}{c^2}U^\a U^\b-\p_b\Delta^{\a\b}\right)\Big]
-\frac{8}{c^2}\n_a\e_a\n_b\e_bU^\a U^\b+8\p_b\p_a^\lab+8\p_a\p_b^\lab$$
$$
-8(\p_a\p_b+\p_a\varpi_b+\p_b\varpi_a)\Delta^{\a\b}-\frac{4}{c^2}\Big[(\n_a\e_a-\p_a)U^\b(\q_b^\a+\h_b \J_b^\a)
+(\n_b\e_b-\p_b)U^\b(\q_a^\a+\h_a \J_a^\a)$$
$$
+(\n_a\e_a-\p_a)U^\a(\q_b^\b+\h_b \J_b^\b)+(\n_b\e_b-\p_b)U^\a(\q_a^\b+\h_a \J_a^\b)\Big]-(m_a^2+m_b^2)c^2\Big[2\n_a\n_bU^\a U^\b$$
$$
+\n_aU^\a \J_b^\b+\n_bU^\b \J_a^\a+\n_bU^\a \J_a^\b+\n_aU^\b \J_b^\a\Big]+2\eta^{\a\b}\left[3(\p_a\p_b+\p_a\varpi_b+\p_b\varpi_a)+\n_a\e_a\n_b\e_\b\right]$$
$$
-2\eta^{\a\b}\Bigg[\left(\frac{m_b}{2m_a}+\frac{m_a}{2m_b}\right)^2
\Big(\n_a\e_a\n_b\e_\b+9(\p_a\p_b+\p_a\varpi_b+\p_b\varpi_a)-3\n_a\e_a(\p_b+\varpi_b)$$
\vskip-1cm\ben\label{40}
-3\n_b\e_b(\p_a+\varpi_a)\Big)\Bigg]\Bigg\}.
\een

\section{Maxwellian iteration procedure}

 In the Maxwellian iteration procedure the equilibrium values of the constitutive quantities   $\p_1^\lab=\p_2^\lab$=0,  $\varpi_1=\varpi_2=0$,  $\q_1^\a=\q_2^\a=0$ and  $\J_1^\a=-\J_2^\a=0$ are inserted on the left-hand sides of the  equations  (\ref{47}) -- (\ref{47c}) and the first iterated values are obtained from the production terms, i.e., from the right-hand side of these equations.

First from (\ref{47}) and (\ref{47b}) we obtain -- by following the above mentioned methodology -- the following algebraic system of equations for the determination of the diffusion $\J_1^\a$ and heat $\q_1^\a$, $\q_2^\a$ fluxes:
\ben
&&\d_1^\a=\frac{\pi\mathcal{B}_1}{2c^2k T}\left[\n\left(\h_1+\h_2-3kT\right)\J_1^\a+\n_2\q_1^\a-n_1\q_2^\a\right],
\een
\ben\no
&&m_1c^2G_1\d_1^\a-\frac{m_1^2c^4}{k T^2}\n_1\left(G_1^2-5\frac{G_1}{\z_1}-1\right)\nabla^\a\mathcal{T}=
-\frac{\pi\mathcal{B}_1}{2c^2 k T}\bigg\{\n_2(5c^2C_{31}-\e_2+3kT)\q_1^\a
\\\no
&&-\n_1(5c^2C_{32}-\e_1+3kT)\q_2^\a-c^2\bigg[\n_1\bigg(C_{52}+C_{62}-\frac{(m_1^2-C_{11})c^2}{6}+\frac{\h_2}{c^2}(\e_1-kT)
\\\no
&&
+\frac{(m_1^2-m_2^2)c^2}{2}\bigg) +\n_2\bigg(C_{51}+C_{61}-\frac{(m_2^2-C_{12})c^2}{6}
+\frac{\h_1}{c^2}(\e_2-kT)
-\frac{(m_1^2-m_2^2)c^2}{2}\bigg) \bigg]\J_1^\a\bigg\}
\\\no
&&
-\frac{\pi\mathcal{B}_2}{8c^2kT}\bigg\{
\bigg[2\n_1(10c^2C_{31}+8kT)+\n_2\bigg(10c^2C_{31}-\bigg(5-\frac{m_1^2}{m_2^2}\bigg)(\e_2-3k T)
+4(\e_2-k T)\bigg)\bigg]\q_1^\a
\\\no
&&+\n_1\bigg[10c^2C_{32}
-\bigg(5-\frac{m_2^2}{m_1^2}\bigg)(\e_1-3k T)+4(\e_1-k T)\bigg]\q_2^\a-2c^2\bigg[2\n_1\bigg(C_{51}+C_{61}-4kT\frac{\h_1}{c^2}
\\\no
&&-\frac{(m_1^2-C_{11})c^2}{6}-m_1^2c^2\bigg)+\n_2\bigg(C_{51}+C_{61}-\frac{(m_2^2-C_{12})c^2}{6}+\bigg(5-\frac{m_1^2}{m_2^2}\bigg)\frac{\e_2-3k T}{2}\frac{\h_1}{c^2}
\\\no
&&-2(\e_2-kT)\frac{\h_1}{c^2}-\frac{(m_1^2+m_2^2)c^2}{2}\bigg)-\n_1\bigg(C_{52}+C_{62}-\frac{(m_1^2-C_{11})c^2}{6}-2(\e_1-kT)\frac{\h_2}{c^2}
\\
&&+\bigg(5-\frac{m_2^2}{m_1^2}\bigg)\frac{\e_1-3k T}{2}\frac{\h_2}{c^2}-\frac{(m_1^2+m_2^2)c^2}{2}\bigg)\bigg]\J_1^\a\bigg\},
\een
\ben\no
&&m_2c^2G_2\d_1^\a+\frac{m_2^2c^4}{k T^2}\n_2\left(G_2^2-5\frac{G_2}{\z_2}-1\right)\nabla^\a\mathcal{T}=
\frac{\pi\mathcal{B}_1}{2c^2 k T}\bigg\{\n_1(5c^2C_{32}-\e_1+3kT)\q_2^\a
\\\no
&&-\n_2(5c^2C_{31}-\e_2+3kT)\q_1^\a+c^2\bigg[\n_1\bigg(C_{52}+C_{62}-\frac{(m_1^2-C_{11})c^2}{6}+\frac{\h_2}{c^2}(\e_1-kT)
\\\no
&&
+\frac{(m_1^2-m_2^2)c^2}{2}\bigg) +\n_2\bigg(C_{51}+C_{61}-\frac{(m_2^2-C_{12})c^2}{6}
+\frac{\h_1}{c^2}(\e_2-kT)
-\frac{(m_1^2-m_2^2)c^2}{2}\bigg) \bigg]\J_1^\a\bigg\}
\\\no
&&
+\frac{\pi\mathcal{B}_2}{8c^2 k T}\bigg\{
\bigg[2\n_2(10c^2C_{32}+8kT)+\n_1\bigg(10c^2C_{32}+4(\e_1-k T)
-\bigg(5-\frac{m_2^2}{m_1^2}\bigg)(\e_1-3k T)\bigg)\bigg]\q_2^\a
\\\no
&&+\n_2\bigg[10c^2C_{31}
+4(\e_2-k T)-\bigg(5-\frac{m_1^2}{m_2^2}\bigg)(\e_2-3k T)\bigg]\q_1^\a+2c^2\bigg[2\n_2\bigg(C_{52}+C_{62}-4kT\frac{\h_2}{c^2}
\\\no
&&-\frac{(m_2^2-C_{12})c^2}{6}-m_2^2c^2\bigg)-\n_2\bigg(C_{51}+C_{61}-\frac{(m_2^2-C_{12})c^2}{6}+\bigg(5-\frac{m_1^2}{m_2^2}\bigg)\frac{\e_2-3k T}{2}\frac{\h_1}{c^2}
\\\no
&&-2(\e_2-kT)\frac{\h_1}{c^2}-\frac{(m_1^2+m_2^2)c^2}{2}\bigg)+\n_1\bigg(C_{52}+C_{62}-\frac{(m_1^2-C_{11})c^2}{6}-2(\e_1-kT)\frac{\h_2}{c^2}
\\
&&+\bigg(5-\frac{m_2^2}{m_1^2}\bigg)\frac{\e_1-3k T}{2}\frac{\h_2}{c^2}-\frac{(m_1^2+m_2^2)c^2}{2}\bigg)\bigg]\J_1^\a\bigg\}.
\een
The solution of the above system of algebraic equations leads to the constitutive equations for $\J_1^\a$, $\q_1^\a$ and $\q_2^\a$:
\ben\lb{nn}
\J_1^\a=-\J_2^\a=\mathcal{D}_{12}\,\d_1^\a+\mathcal{D}_T \nabla^\a\mathcal{T},\quad\q_1^\a=\lambda_1\nabla^\a\mathcal{T}+\mathcal{D}_1\,\d_1^\a,\quad \q_2^\a=\lambda_2\nabla^\a\mathcal{T}+\mathcal{D}_2\,\d_1^\a,
\een
which are linear functions of the relativistic temperature gradient
\ben
\nabla^\a\mathcal{T}\equiv \nabla^\a T-\frac{T}{\n\h}\nabla^\a \p,
\een
and of the generalized diffusion force
\ben
\d_1^\a\equiv \frac{1}{k T}\left(\nabla^\a \p_1-\frac{\n_1\h_1}{\n\h}\nabla^\a \p\right),\qquad\hbox{with}\qquad \d_2^\a=-\d_1^\a.
\een
Equation (\ref{nn}a) represents generalized Fick's law, while (\ref{nn}b,c) generalized Fourier's law. Here $\lambda_1,$ $\lambda_2,$ $\mathcal{D}_{12},$ $\mathcal{D}_T,$ $\mathcal{D}_1,$  $\mathcal{D}_2$ are scalar coefficients.

The pressure deviator tensors $\p_1^\lab, \p_2^\lab$ are obtained from the algebraic system
\ben\no
&&\frac{c^2}{3}\n_1\left(m_1^2- C_{11}\right) \nabla^{\langle\a}U^{\b\rangle}
=-\frac{\pi}{2c^2}\mathcal{B}_1\bigg\{\n_2\left(c^2C_{41}+\e_2-3kT\right)\p_1^\lab -\n_1\big(c^2C_{42}+\e_1
\\\no
&&-3kT\big)\p_2^\lab\bigg\}-\frac{\pi}{8c^2}\mathcal{B}_2\bigg\{\bigg[2\n_1\left(2c^2C_{41}+4\e_1-4k T\right)+\n_2\bigg(2c^2C_{41}+\bigg(5-\frac{m_1^2}{m_2^2}\bigg)(\e_2
\\
&&-3kT)+8kT
\bigg)\bigg]\p_1^\lab+\n_1\bigg[2c^2C_{42}+\left(5-\frac{m_2^2}{m_1^2}\right)\left(\e_1-3kT\right)
+8kT\bigg]\p_2^\lab\bigg\},
\een
\ben\no
&&\frac{c^2}{3}\n_2\left(m_2^2- C_{12}\right) \nabla^{\langle\a}U^{\b\rangle}
=-\frac{\pi}{2c^2}\mathcal{B}_1\bigg\{\n_1\left(c^2C_{42}+\e_1-3kT\right)\p_2^\lab -\n_2\big(c^2C_{41}+\e_2
\\\no
&&-3kT\big)\p_1^\lab\bigg\}-\frac{\pi}{8c^2}\mathcal{B}_2\bigg\{\bigg[2\n_2\left(2c^2C_{42}+4\e_2-4k T\right)+\n_1\bigg(2c^2C_{42}+\bigg(5-\frac{m_2^2}{m_1^2}\bigg)(\e_1
\\
&&-3kT)
+8kT\bigg)\bigg]\p_2^\lab+\n_2\bigg[2c^2C_{41}+\left(5-\frac{m_1^2}{m_2^2}\right)\left(\e_2-3kT\right)
+8kT\bigg]\p_1^\lab\bigg\},
\een
that follow from (\ref{47c}) by using the Maxwellian iteration procedure. From the above system of algebraic equations we obtain the constitutive equations for the pressure deviator tensors as linear functions of the traceless part of the velocity gradient $\nabla^{\langle\a}U^{\b\rangle}$, namely.
\ben\lb{mm1}
\p_1^\lab=2\mu_1 \nabla^{\langle\a}U^{\b\rangle},\qquad \p_2^\lab=2\mu_2 \nabla^{\langle\a}U^{\b\rangle},
\een
where $\mu_1, \mu_2$ are scalar coefficients. The traceless part of the gradient of velocity is defined in the same way as the tensor $\Upsilon^{\a\b}$ in (\ref{ups}).

Finally, the Maxwellian iteration procedure when applied to (\ref{47a}) leads to the following system of algebraic equations for the non-equilibrium pressures $\varpi_1, \varpi_2$:
\ben\no
&&\n_1\left[\frac{C_{11}-m_1^2}{3}+\frac{3km_1^2}{c_v}\left(G_1^2-6\frac{G_1}{\z_1}-1\right)\right] \nabla_\a U^\a
=-\frac{\pi}{4c^4}\mathcal{B}_1\bigg\{\n_2\left(c^4C_{21}+6\e_2\right)\varpi_1
\\\no
&&-\n_1\left(c^4C_{22}+6\e_1\right)\varpi_2\bigg\}-\frac{\pi}{8c^4}\mathcal{B}_2\bigg\{\underline{\frac{3\n_1\n_2}{2}\bigg[\left(m_1^2-m_2^2\right)\left(\e_1\e_2-3 k^2 T^2\right)-3kT\left(m_1^2\e_1-m_2^2\e_2\right)}
\\\no
&&\underline{-kT\left(m_1^2\e_2-m_2^2\e_1\right)\bigg]\frac{m_2^2-m_1^2}{m_1^2m_2^2}}
+2\n_1\left[c^4C_{21}-6\left(\e_1+2k T\right)\right]\varpi_1+\n_2\bigg[c^4C_{21}+6k T
\\\no
&&-3\left(5-\frac{m_2^2}{m_1^2}\right)\e_2+6\left(\frac{m_1^2+m_2^2}{2m_1m_2}\right)^2
\left(\e_2-3k T\right)\bigg]\varpi_1+\n_1\bigg[c^4C_{22}+6k T-3\left(5-\frac{m_1^2}{m_2^2}\right)\e_1
\\\lb{nn1}
&&+6\left(\frac{m_1^2+m_2^2}{2m_1m_2}\right)^2\left(\e_1-3k T\right)\bigg]\varpi_2\bigg\},
\een
\ben\no
&&\n_2\left[\frac{C_{12}-m_2^2}{3}+\frac{3km_2^2}{c_v}\left(G_2^2-6\frac{G_2}{\z_2}-1\right)\right] \nabla_\a U^\a
=-\frac{\pi}{4c^4}\mathcal{B}_1\bigg\{\n_1\left(c^4C_{22}+6\e_1\right)\varpi_2
\\\no
&& -\n_2\left(c^4C_{21}+6\e_2\right)\varpi_1\bigg\}-\frac{\pi}{8c^4}\mathcal{B}_2\bigg\{\underline{\frac{3\n_1\n_2}{2}\bigg[\left(m_2^2-m_1^2\right)\left(\e_1\e_2-3 k^2 T^2\right)-3kT\left(m_2^2\e_2-m_1^2\e_1\right)}
\\\no
&&\underline{-kT\left(m_2^2\e_1-m_1^2\e_2\right)\bigg]\frac{m_1^2-m_2^2}{m_1^2m_2^2}}+2\n_2\left[c^4C_{22}-6\left(\e_2+2k T\right)\right]\varpi_2+\n_2\bigg[c^4C_{21}+6k T
\\\no
&&-3\left(5-\frac{m_2^2}{m_1^2}\right)\e_2+6\left(\frac{m_1^2+m_2^2}{2m_1m_2}\right)^2\left(\e_2-3k T\right)\bigg]\varpi_1+\n_1\bigg[c^4C_{22}-3\left(5-\frac{m_1^2}{m_2^2}\right)\e_1+6k T
\\\lb{nn2}
&&+6\left(\frac{m_1^2+m_2^2}{2m_1m_2}\right)^2\left(\e_1-3k T\right)\bigg]\varpi_2\bigg\}.
\een
We note here that the above underlined terms are of order $\epsilon^2$ which must be neglected due to the hypothesis that the masses of the particles of the constituents  are not too disparate, i.e.,  $m_2=m_1(1+\epsilon)$ with $\epsilon$ being a small quantity. Hence we obtain from (\ref{nn1}) and (\ref{nn2}) the linear relationships between the non-equilibrium pressures and the velocity divergent $\nabla_\a U^\a$:
\ben\lb{mm2}
\varpi_1=-\eta_1 \nabla_\a U^\a,\qquad \varpi_2=-\eta_2\nabla_\a U^\a,
\een
where $\eta_1, \eta_2$ are scalar coefficients.

Equations (\ref{mm1}) and (\ref{mm2}) represent the constitutive equations of a relativistic Newtonian fluid, also known as  Navier-Stokes law.

\section{Generalized diffusion force and relativistic temperature gradient in the $\epsilon$-order}

In the $\epsilon$-order the generalized diffusion force and the relativistic temperature gradient read
\ben
&&\d_1^\a=\x_2\nabla^\a\n_1-\x_1\nabla^\a\n_2+\epsilon\,\x_1\x_2\,\frac{G_1^2\z_1-\z_1-4G_1}{G_1}\left[\nabla^\a\n_1+\nabla^\a\n_2+\frac{\n}{T}\nabla^\a T\right],
\\\no
&&\nabla^\a\mathcal{T}=\nabla^\a T-\frac{T}{\n\h}\nabla^\a \p=\nabla^\a T-\frac{1}{\z_1G_1}\left(1-\epsilon\,\x_2\,\frac{G_1^2\z_1-\z_1-4G_1}{G_1}\right)
\\&&\qquad\times
\left[\nabla^\a T+\frac{T}{\n}\nabla^\a \n_1+\frac{T}{\n}\nabla^\a \n_2\right].
\een


 \end{document}